\documentclass[preprint,showpacs,preprintnumbers,amsmath,amssymb, superscriptaddress, letterpaper, prb]{revtex4}
\usepackage{graphicx}
\usepackage{dcolumn}
\usepackage{bm}

\begin{document}

\preprint{}

\title{Classical Disordered Ground States: Super-Ideal Gases, and Stealth and Equi-Luminous Materials}
\author{Robert D. Batten}
\affiliation{ Department of Chemical Engineering, Princeton University, Princeton, NJ 08544, USA}
 
\author{Frank H. Stillinger}
\affiliation{Department of Chemistry, Princeton University, Princeton, NJ, 08544 USA}

\author{Salvatore Torquato}
\affiliation{Department of Chemistry, Princeton University, Princeton, NJ, 08544 USA}
\affiliation{Princeton Materials Institute, Princeton University, Princeton, NJ 08544, USA}
\affiliation{Program in Applied and Computational Mathematics, Princeton University, Princeton, NJ 08544, USA}
\affiliation{Princeton Center for Theoretical Physics, Princeton University, Princeton, NJ 08544, USA}
\affiliation{School of Natural Sciences, Institute for Advanced Study, 
Princeton, NJ, 08544 USA}

\altaffiliation{Corresponding author}
\email{torquato@electron.princeton.edu}
\date{\today}

\begin{abstract}
Using a collective coordinate numerical optimization procedure, we construct ground-state configurations of interacting particle systems in various space dimensions so that the scattering of radiation exactly matches a prescribed pattern for a set of wave vectors.  We show that the 
constructed ground states are, counterintuitively, disordered (i.e., possess no long-range order) in the infinite-volume limit. We focus on three classes of configurations with unique radiation scattering characteristics: (i)``stealth'' materials, which are transparent to incident radiation at certain wavelengths; (ii)``super-ideal'' gases, which scatter radiation identically to that of an ensemble of ideal gas configurations for a selected set of wave vectors; and (iii)``equi-luminous'' materials, which scatter radiation equally intensely for a selected set of wave vectors.  We find that ground-state configurations have an increased tendency to contain clusters of particles as one increases the prescribed luminosity. Limitations and consequences of this procedure are detailed.
\end{abstract}

\pacs{}
\maketitle

\section{Introduction}
\label{sec:intro} 
A fundamental problem of statistical mechanics is the determination and understanding of classical ground states of many-particle systems - the zero-temperature particle arrangement that minimizes potential energy per particle.  Although perfect crystalline (periodic) structures are often energetically favorable among all configurations, there is incomplete mathematical and intuitive understanding of the formation of order at low temperature, \cite{radin1987lta} introducing the possibility of disordered ground states. A disordered many-particle system is one that lacks long-range order. More precisely, disordered systems have a pair correlation function $g_2({\bf r})$ (defined below) that decays to unity faster than $|{\bf r}|^{-d-\epsilon}$, for spatial dimension $d$ and some positive $\epsilon$, in the infinite-volume limit. \cite{torquato2006ncl} Recently, a collective coordinate approach has been used to identify certain possibly disordered ground states. \cite{fan1991ccd, uche2004ccd, uche2006ccc}  Although the previous studies are suggestive of the relation between disordered ground states and collective coordinates for finite systems, a systematic investigation of disordered ground states, including whether they exist in the infinite-volume limit, is not yet available.
 
In this paper, we use an ``inverse'' approach to construct classical disordered ground states with precisely tuned wave scattering characteristics via the aforementioned collective coordinate procedure.  In a recent example of a ``forward'' problem, the scattering from glass ceramics with nanometer-sized crystals was likened to that of random sequential adsorption (RSA) of hard spheres, \cite{mattarelli2007ugc} a well-known, disordered many-body configuration. \cite{torquato2006rsa}  These ceramics are of interest in photonics applications because they are mechanically rigid and nearly suppress all scattering at long wavelengths. \cite{joannopoulos1995pcm} 

By contrast, our method utilizes an inverse approach: we prescribe scattering characteristics ({\it e.g.,} absolute transparency) and construct many-body configurations that give rise to these targeted characteristics.  Potential applications include designing ground-state materials as radiation filters or scatterers, and materials transparent to specific wavelengths of radiation, among others.  We apply our methodology initially for structureless (i.e., point) particles. However, it could be generalized to structured particles, colloids, or as bodies using the appropriate structure factors for finite-sized particles as is done for random media. \cite{torquato2002rhm}

Since previous studies utilized small periodic systems, \cite{fan1991ccd, uche2004ccd, uche2006ccc} we first establish that systematically increasing the system size has no effect on the degree of disorder.  Extrapolation from these results indicates that the constructed configurations remain disordered in the infinite-volume limit;
a seemingly counterintuitive proposition.
We then construct disordered ground states with special scattering properties: ``stealth materials,'' ``super-ideal gases,'' and ``equi-luminous materials.''  

We use the term ``stealth'' materials to refer to many-particle configurations that completely suppress scattering of incident radiation for a set of wave vectors, and thus, are transparent at these wavelengths. \cite{footnote}  Periodic (i.e., crystalline) configurations are, by definition, ``stealthy'' since they suppress scattering for all wavelengths except those associated with Bragg scattering. However, we construct disordered stealth configurations that prevent scattering only at prescribed wavelengths with no restrictions on any other wavelengths. 

We define a ``super-ideal'' gas as a single many-particle configuration whose scattering {\it exactly} matches that of an ensemble of ideal gas configurations, or Poisson point distributions, for a set of wave vectors.  A super-ideal gas has a structure factor that is identically unity for specified wave vectors, and thus, this single configuration would be impossible to differentiate from an ensemble of ideal gas configurations for the specified wave vectors.  

We define an ``equi-luminous'' material to be a system whose scattering is constant for a set of wavelengths. Although stealth materials and super-ideal gases are subsets of equi-luminescent materials, we use this term to refer to materials that scatter radiation more intensely relative to an ideal gas.  These materials that scatter radiation much more intensely than an ideal gas for a set of wave vectors have enhanced density fluctuations and show local clustering similar to polymers and aggregating colloids. \cite{schaefer1989pfa} Typically, scattering experiments on these systems are used to shed light on the characteristic length scales of a system.\cite{schaefer1989pfa, martin1987sf, teixeira1988sas}  With our inverse procedure, we impose the degree of clustering by tuning the scattering characteristics for certain wavelengths.   

Upon generating ensembles of ground-state configurations for each class of materials described above, we characterize local order of each ensemble.  We place emphasis on pair information in real space via the pair correlation function $g_2({\bf r})$ and in reciprocal space through the structure factor $S({\bf k})$ as these functions are experimentally accessible and used widely in many-body theories. \cite{torquato2002rhm, ashcroft1976ssp, feynman1998sm}

The pair correlation function $g_2({\bf r})$ is the normalized two-particle probability density function $\rho_2({\bf r})$ and is proportional to the probability of observing a particle center at ${\bf r}$ relative to a particle at the origin.\cite{torquato2002rhm}  For a statistically homogeneous and isotropic medium, the pair correlation function $g_2({\bf r})$ depends only on the magnitude of $r\equiv|{\bf r}|$, and is commonly referred to as the radial distribution function $g_2(r)$, which henceforth is the designation used in this paper.

The structure factor $S({\bf k})$ is proportional to the intensity of scattering of incident radiation from a configuration of $N$ particles and is defined as 
\begin{equation}
\label{eq:sk}
S({\bf k}) = \frac{|\rho({\bf k})|^2}{N}, 
\end{equation}
where $\rho({\bf k})$ are the collective coordinates and ${\bf k}$ are the wave vectors associated with the system volume and boundary conditions. Collective coordinates $\rho({\bf k})$ are the Fourier coefficients in the expansion of the density field:
\begin{equation}
\label{eq:rhok}
\rho({\bf k }) =  \sum_{j=1}^{N}  \exp(i{\bf k \cdot r}_j),
\end{equation}
where ${\bf r}_j$ denotes the location of particle {\it j}. When $S({\bf k})$ depends only on the the magnitude of $k\equiv|{\bf k}|$, the structure factor $S(k)$ is related to the Fourier transformation of $g_2(r)-1$, ignoring the forward scattering associated with ${\bf k}={\bf 0}$,
\begin{equation}
\label{eq:sktogr}
S(k) = 1 + \rho \int \exp(i {\bf k \cdot r}) \left[g_2(r)-1\right]d{\bf r},
\end{equation}
where $\rho$ is the number density. For highly ordered systems, both $g_2(r)$ and $S(k)$ contain a series of $\delta$-functions or peaks at large $r$ and $k$, indicating strong correlations at the associated pair distance.  In configurations without long-range order, both $g_2(r)$ and $S(k)$ approach unity at large $r$ and $k$. 

Several inverse methods have sought to construct systems using pair information in real space, particularly in addressing the question of pair correlation function ``realizability,''\cite{torquato2002rhm, torquato2002csr, costin2004cpd,  stillinger2004pcf, stillinger2005rii, torquato2006ncl} which asks whether a given pair correlation function, at number density $\rho$, can be realized by spatial arrangements of particles. Typically, in real space, a ``target'' radial distribution function is chosen and many-particle configurations are found that best match the target $g_2(r)$.  Stochastic optimization techniques have been a popular reconstruction method in finding spatial arrangements of particles that best approximate a target correlation function.\cite{yeong1998rrm, cule1999grm, sheehan2001gms, crawford2003acf, uche2006rpc}  

In contrast to these real-space methods, we target pair information in reciprocal space to construct configurations whose structure factor {\it exactly} matches the candidate structure factor for a set of wavelengths.  In addition, our procedure guarantees that the resulting configuration is a ground-state structure for a class of potential functions.

The remainder of this paper is as follows.   Section \ref{sec:background} presents background on disordered ground states and motivates our choices for candidate structure factors, while Section \ref{sec:methods} outlines the numerical procedure.  Structure factors, radial distribution functions, and representative particle patterns for stealth materials, super-ideal gases, and equi-luminous materials are found in Secs.\ \ref{sec:stealth}, \ref{sec:superideal}, and \ref{sec:lumin}.  Lastly, general conclusions and discussion relevant to classical disordered ground states and this procedure are found in Sec.\ \ref{sec:conclusions}. Appendix \ref{sec:energy} compares minimization algorithms and analyzes the energy landscapes associated with our potential functions.

\section{Stealth Materials, Super-Ideal Gases, and Equi-Luminous Materials}
\label{sec:background} 

Physical intuition and experimental facts suggest that in the zero-temperature limit, classical systems of interacting particles adopt a periodic structure to minimize potential energy.  The ``crystal problem'' has attempted to determine the fundamental mechanism that forces particles into ordered states, but the existence of these mechanisms has yet to be fully understood. \cite{radin1987lta}  The notion of disordered ground states is particularly mysterious because of the lack of symmetry, lack of long-range order, and degeneracy of ground-state configurations.

The characterization of order in solid phases in the low-temperature limit has been well studied.  In addressing the crystal problem, it has been suggested that nonanalyticity of thermodynamic functions may yield ``turbulent,'' or nonperiodic, Gibbs states at positive temperature. \cite{ruelle1981mtf}  As a consequence, a turbulent crystal, characterized by fuzzy diffraction peaks, is possible as a nonperiodic solid phase, in addition to periodic and quasiperiodic structures. \cite{ruelle1982tce}  Turbulent crystals have been examined previously \cite{ruelle1982tce, yukalov2001ptc} and evidence has been presented that at low temperature, equilibrium states may contain disorder. \cite{butz1984ers} Theoretical work has created classical lattice models with short-range interactions whose ground states contain the property of disorder. \cite{radin1991dgs}  In addition, a simple gradient model was used to develop a disordered state, a labyrinth, within a pattern forming system.  Although this state was ultimately excluded from being a ground state, the authors were unable to exclude other models as potentially yielding disordered ground states. \cite{leberre2002ecc}  Despite significant research attention, understanding of the attainability of disordered configurations as classical grounds states is incomplete.  

We choose to limit this investigation to the study of stealth materials, super-ideal gases, and equi-luminous materials based on the potential applications and fundamental interest.  Constructing systems with transparency at specified wavelengths, including wavelengths outside of the low-$k$ region, is the primary motivation. A stealth material has a structure factor that is exactly zero for some set of wavelengths. In the low-$k$ limit, several disordered systems nearly suppress all scattering, though most are not ground states. For example, when crystallizing hard colloidal spheres, stacking faults of close-packed layers create deviation from perfect crystallinity.  The structure factor is nearly zero in the low-$k$ region followed by a strong Bragg peak and a diffuse peak before decaying to unity. \cite{pusey1989sch}  

The examples of stealth configurations can also be described as ``hyperuniform.''  Hyperuniform systems have the property that 
\begin{equation}
\lim_{k\rightarrow 0} S(k) = 0,
\end{equation}
i.e., infinite-wavelength density fluctuations vanish.\cite{torquato2003ldf} Hyperuniform point patterns arise in the structure of the early Universe,\cite{gabrielli2002glu} maximally random jammed packings,\cite{Ka02,torquato2003ldf, donev2005udf} certain tilings of space,\cite{gabrielli2004vav} and the ground-state configurations of certain repulsively interacting particle systems.\cite{torquato2003ldf} 

The structure factor for a super-ideal gas is exactly unity for a set of wave vectors and unconstrained for the remaining wave vectors. We narrow our study to constraining wave vectors in the small $k$ region. We choose super-ideal gases based on interest in Poisson point distributions.  The Poisson point process has the simplest candidate $S(k)$ and $g_2(r)$, both being exactly unity for all $k$ and $r$, respectively.  For a single finite configuration, the structure factor exhibits random fluctuations about unity. But in a super-ideal gas, we constrain a set of wave vectors so that $S(k)$ is exactly unity and, for all wave vectors outside of this unconstrained set, the intuitive expectation is that the structure factor would ensemble-average to unity. In fact, this does not necessarily happen and an interesting alternative behavior arises, discussed in detail below. 

``Equi-luminous'' describes materials that scatter light equally intensely for a set of wave vectors. The structure factor for this class of materials is simply a constant for a set of wave vectors.  Subsets of equi-luminous materials include super-ideal gases ($S(k)=1$) and stealth materials ($S(k)=0$). Here, we focus on materials whose structure factor is a constant greater than unity, as these systems show strong local clustering and intense scattering relative to an ideal gas.

\section{Numerical Procedure}
\label{sec:methods}
The numerical optimization procedure follows that of Uche, Stillinger and Torquato \cite{uche2006ccc} used to tailor the small $k$ behavior of the structure factor. The structure factor $S(k)$ and collective coordinates $\rho({\bf k})$, defined in Eqs.\ (\ref{eq:sk}) and (\ref{eq:rhok}), are related to the quantity $C({\bf k})$,
\begin{equation}
\label{eq:sktock}
S({\bf k}) = 1 + \frac{2}{N}C({\bf k}), 
\end{equation}
where
\begin{equation}
\label{eq:ck}
C({\bf k}) = \sum_{j=1}^{N-1} \sum_{i=j+1}^{N} \cos\left[ {\bf k} \cdot \left( {\bf r}_j - {\bf r}_i \right) \right].
\end{equation}
For a system interacting via a pair potential $v({\bf r}_i-{\bf r}_j)$, the total potential energy can be written in terms of $C({\bf k})$, 
\begin{eqnarray}
\Phi &=&\sum_i \sum_j v({\bf r}_i-{\bf r}_j) \\
\label{eq:phi}
     &=& \Omega^{-1} \sum_{{\bf k}} V({\bf k})C({\bf k}), 
\end{eqnarray}
where $\Omega$ is the system volume and $V({\bf k})$ is the Fourier transform of the pair potential function
\begin{equation}
V({\bf k}) = \int_\Omega d{\bf r} v({\bf r}) \exp(i {\bf k} \cdot {\bf r}).
\end{equation}
For a region of space with dimensions $L_x$, $L_x \times L_y$, or $L_x \times L_y \times L_z$ in one, two, or three dimensions, subject to periodic boundary conditions, the infinite set of corresponding wave vectors has components
\begin{equation}
k_\gamma = \frac{2\pi n_\gamma}{L_\gamma},
\end{equation}
where $n_\gamma$ are positive or negative integers, or zero and $\gamma$=$x,y,z$ as needed.  For example, in three dimensions, the set of wave vectors are 
\begin{equation}
{\bf k} = \left( \frac{2\pi n_x}{L_x}, \frac{2\pi n_y}{L_y}, \frac{2\pi n_z}{L_z} \right).
\end{equation}

We introduce a square mound $V({\bf k})$ that is a positive constant $V_0$ for all {\bf k} $\in {\bf Q}$, where ${\bf Q}$ is the set of wave vectors such that $0 < |{\bf k }| \le K$, and zero for all other {\bf k}.  In the infinite-volume limit, this corresponds to a system of particles interacting via a real-space pair potential function that is bounded, damped, and oscillating about zero at large $r$.\cite{fan1991ccd, uche2004ccd}  For a cutoff radius $K$, there are $2M(K)$ wave vectors in the set ${\bf Q}$, where $M(K)$ is the number of independently constrained collective coordinates.  That is, constraining $C({\bf k})$ implicitly constrains $C(-{\bf k})$ due to the relation 
\begin{equation}
C({\bf k}) = C(-{\bf k}).
\end{equation}
For a system of $N$ particles in $d$ dimensions, there are $dN$ total degrees of freedom. We introduce the dimensionless parameter $\chi$ to conveniently represent the ratio of the number of constrained degrees of freedom relative to the total number of degrees of freedom
\begin{equation}
\chi = \frac{M(K)}{dN}. 
\end{equation}
The global minimum of the potential energy defined in Eq.\ (\ref{eq:phi}) has the value of 
\begin{equation}
\label{eq:min}
\min_{{\bf r}_1\cdots{\bf r}_N}(\Phi) = -\left( \frac{N}{2}\right) \sum_{{\bf k} \in {\bf Q}}V_0,  
\end{equation} 
if and only if there exist particle configurations that satisfy all of the imposed constraints, which necessarily occurs for $\chi\leq 1$.  Minimizing Eq.\ (\ref{eq:phi}) to its global minimum, for $\chi\leq 1$, yields ground-state configurations that are stealthy for all {\bf k} $\in {\bf Q}$.  

To target a specific form of the structure factor to certain nonzero values, such as $S(k)$ = 1, we introduce a second nonnegative objective function,
\begin{equation}
\label{eq:phitilde}
\Phi = \sum_{{\bf k}\in Q} V({\bf k}) \left[C({\bf k}) -C_0({\bf k}) \right]^2,  
\end{equation}
where $C_0({\bf k})$ is associated with the target structure factor by Eq.\ (\ref{eq:sktock}).  If Eq.\ (\ref{eq:phitilde}) is taken to be the potential energy of an $N$-body system, then two-, three-, and four- body interactions are present.\cite{uche2006ccc}  Equation (\ref{eq:phitilde}) has a global minimum of zero, for $\chi \leq 1$, if and only if there exist configurations that satisfy all of the imposed constraints.  Minimizing Eq.\ (\ref{eq:phitilde}) is used to construct super-ideal gases and equi-luminous materials as ground-state configurations.

Three algorithms have been employed previously for minimizing Eqs.\ (\ref{eq:phi}) and (\ref{eq:phitilde}): steepest descent,\cite{fan1991ccd} conjugate gradient,\cite{uche2004ccd} and MINOP.\cite{uche2006ccc}  Steepest descent and conjugate gradient methods are line search methods that differ only in their choice of search directions. \cite{press1992nrc}  The MINOP algorithm is a trust-region method.  When far from the solution, the program chooses a gradient direction, but when close to the solution, it chooses a quasi-Newton direction.\cite{dennis1979tnu, kaufman1999rsq}  Upon each iteration, the program makes an appropriate update to approximate the Hessian.\cite{kaufman1999rsq}  

We find that that neither the conjugate gradient method nor MINOP algorithm significantly biases any subset of ground-state configurations.  The resulting configurations are visually similar, and the ensemble-averaged radial distribution function and structure factor produced by both methods have similar features. We chose the MINOP algorithm because it has been demonstrated to be better suited to the collective coordinate procedure than the conjugate gradient method. \cite{uche2006ccc}  We refer the reader to Appendix \ref{sec:energy} for characterization of the energy landscape and comparison between line search methods and MINOP.    

Three sets of initial conditions were considered: random placement of particles (Poisson distributions), random sequential addition (RSA), and perturbed lattices (integer, triangular, and face centered cubic in one, two, and three dimensions respectively). For an RSA process, particles are assigned a diameter and randomly and irreversibly placed in space such that particles are not overlapping. \cite{torquato2002rhm} At sufficiently high $\chi$, usually $\chi \ge$ 0.6, the constructed ground-state systems apparently lose all memory of their initial configurations. The analyses presented in the following sections will be those of random initial conditions.  In some cases at large $\chi$, a global minimum is not found.  For the results discussed here, Eqs.\ (\ref{eq:phi}) and (\ref{eq:phitilde}) were minimized to within $10^{-17}$ of their respective minimum value.  All other trials were excluded from the analysis. 

The region of space occupied by the $N$ particles was limited to a line in one dimension, a square in two dimensions, and a cube in three dimensions, with periodic boundary conditions.  For stealth materials, particular attention was paid to the choice of $N$ for two and three dimensions.  Minimizing Eq.\ (\ref{eq:phi}) for large $\chi$ is known to yield crystalline ground states.\cite{fan1991ccd, uche2004ccd, uche2006ccc}  We choose to be consistent with previous studies.  In two dimensions, $N$ was chosen as a product of the integers $2pq$, and $p/q$ is a rational approximation to $3^{1/2}$ so that all particles could be placed in a triangular lattice configuration without substantial deformation.  In three dimensions, $N$ was usually chosen so that $N=4s^3$, where $s$ is an integer, so that the particles could be placed in a face centered cubic lattice without deformation. In minimizing Eq.\ (\ref{eq:phitilde}), $N$ occasionally was assigned other values.

\section{Results for Stealth Materials}
\label{sec:stealth}

\subsection{Infinite-Volume Limit}
Previous work suggesting the existence of disordered ground states utilized small simulation boxes containing up to several hundred particles. \cite{fan1991ccd, uche2004ccd, uche2006ccc}  Our goal here is to show that constructed systems continue to show no long-range order in the infinite-volume limit. For $d$=2 and 3, systems containing up to several thousand particles were constructed by minimizing Eq.\ (\ref{eq:phi}) for small $\chi$ values.  We find that ground-state configurations differing only by $N$, with $\rho$ and $\chi$ fixed, are disordered and exhibit the same local structure.

Figures \ref{fig:stealth3DinfSK} and \ref{fig:stealth3DinfGR} demonstrate the behavior of $S(k)$ and $g_2(r)$ for stealth materials constrained at $\chi$=0.05.  The structure factors for systems only differing in $N$ have identical characteristics.  The structure factor is exactly zero for constrained wave vectors and subsequently peaks above and fluctuates about unity.  The averaged $S(k)$ initially peaks to a value of 1.10 and decays rapidly to unity, a feature that is more apparent for a large ensemble of configurations.

The radial distribution function remains essentially invariant as the number of particles in the simulation box increases from 108 to 6912, with $\chi$ and $\rho$ fixed, as shown in Figure \ref{fig:stealth3DinfGR}.  For a single realization, a system containing 108 particles has the same shape of $g_2(r)$ as that of the larger system but shows significant statistical noise. In the figure, we ensemble-average the results for the smaller system to make clear the structural similarities.  In the smaller system, the large-$r$ behavior is unavailable due to the minimum image convention of the periodic box. Thus, the figure only displays local structure, which is clearly disordered.  In both cases, $g_2(r)$ dips slightly below unity for small $r$ and quickly approaches and oscillates about unity with a diminishing amplitude.

\subsection{Effect of Increasing Constraints}
It was previously reported that minimizing collective density variables for sufficiently high $\chi$ induces crystallization, \cite{uche2006ccc, fan1991ccd, uche2004ccd} therefore, for disordered, stealth ground states, we have minimized Eq.\ (\ref{eq:phi}), focusing on the low $\chi$ regime.  

For $d=1$, crystallization occurs for $\chi >$ 0.5.\cite{fan1991ccd} For most values of $\chi$ below the crystallization threshold, the structure factor is zero for all constrained wave vectors, it then peaks above unity immediately outside $K$ and decays toward one.  As $\chi$ is increased, the height of the peak decreases and at $\chi$ higher than 0.30, a second peak forms.  Below the crystallization value, $S(k)$ dampens to unity for $k$ larger than $K$.

For $d=2$, three $\chi$ regimes have been reported: disordered for $\chi<$0.57, ``wavy crystalline'' for $0.57\leq\chi<0.77$, and crystalline for $\chi\geq 0.77$. \cite{uche2004ccd}  We choose to investigate well below these ordered regions.  In the ensemble-averaged structure factor, a peak forms in $S(k)$ immediately beyond $K$ and decays toward unity. As $\chi$ increases, the magnitude of the peak increases, in contrast to $d=1$, and at the wavy crystalline threshold, several peaks begin to form beyond $K$. The height of the peak is density dependent, however, $S(k)$ generally has a maximum between $1.1-2.0$ in the disordered region. 

For $d=3$, the transition from disordered to crystalline regimes was identified previously to occur at $\chi$ near 0.5.\cite{uche2006ccc} Constraining $\chi$ below 0.45, $S(k)$ peaks immediately beyond $K$ and smoothly decays to unity, but at $\chi$ = 0.45467, $S(k)$ smoothly oscillates about unity.  The magnitude of the peak is generally smaller than for systems similarly constrained in lower dimensions. In Figure \ref{fig:stealth3Dsk}, we compare the structure factor for several $\chi$ values for 500 particles in a unit cube. We include a nearly crystalline system of 500 particles constrained at $\chi$=0.54867.  The order of the system is apparent by the series of sharp peaks in $S(k)$ that persist at large $k$.

We find that particles have a repellent core that increases in strength with increasing $\chi$. Figure  \ref{fig:stealth3Dgr} demonstrates the repellent core effect via the radial distribution function associated with the 500 particle system described above.  For this particular system, at $\chi$ = 0.45467, an exclusion region develops where $g_2(r)$ is exactly zero for a region near the origin.  At $\chi$ = 0.54867, the peaks demonstrate crystallinity.  
 
Increasing $\chi$ tends to increase the the net repulsion of the potential, which is clearly observed in particle patterns.  Since differentiating between disordered stealth systems is most instructive in two dimensions, we present particle patterns in this dimension only.  Figure \ref{fig:config2dstealth} compares particle patterns of 168 particles with Eq.\ (\ref{eq:phi}) constrained for small $\chi$.  The circular window in the figure represents the length scale of the wavelength associated with $K$. At the lowest $\chi$ considered, the particles do not appear to have any spatial correlation. At higher $\chi$ values, particles develop an exclusion shell about their center but do not have any long-range order.

\subsection{Stealth Materials Spherically Constrained in $k$ Space}

The entirety of disordered stealth materials studied have involved constraining wave vectors in a spherical shell near the origin.  Since this procedure is capable of constraining wave vectors of choice, we have constructed configurations in which two disconnected, concentric regions of wave vectors are constrained, one near the origin and a second further from the origin.  

We define parameters $K_0< K_1 < K_2 <K_3$ as magnitudes of limits for constrained wave vectors.  In this class of stealth configurations, we constrain collective coordinates so that $S(k)$ is zero for all wave vectors in two spherical shells about the origin in reciprocal space.  Specifically, $S(k)$ is zero for all $K_0 < |{\bf k}| \le K_1$ and all $K_2 < |{\bf k}| \le K_3$.  The region $K_1 < |{\bf k}| \le K_2$ is defined as the intermediate region, where the structure factor can be free to fluctuate or to be controlled.  Figure \ref{fig:stealth3Dsplit} shows the location of parameters.  These systems are constructed by introducing a $V({\bf k})$ that contains two square mounds.

Spherically constrained stealth configurations were constructed in two and three dimensions.  We present ensemble-averaged radial distribution functions and structure factors for $d=3$ and particle patterns for $d=2$ as these provide the clearest representation of the general trends.  

The peaking phenomenon in $S(k)$, as originally observed in simple stealth materials, is evident both immediately beyond $K_1$ and beyond $K_3$.  The structure factor increases above unity slightly in the intermediate region but peaks and decays rapidly to unity beyond $K_3$.  This is seen for all test cases and it is not immediately clear if this phenomenon persists for large separations of constrained regions (i.e., $K_2-K_1$). Figure \ref{fig:stealth3Dsplit} shows the ensemble-averaged radial distribution function and structure factor for configurations of 500 particles constrained so that $S(k)$ = 0 for  $0<|{\bf k}| \le 8.8\pi$ and $13\pi < |{\bf k}| \le 14.8\pi$.

An important feature of this procedure is the ability to suppress scattering for wave vectors that are normally Bragg peaks in a crystalline material.  With a 500 particle, three-dimensional system, minimizing Eq.\ (\ref{eq:phi}) for a single square mound $V({\bf k})$ for $\chi$ = 0.54867 creates a crystalline ground state where the first Bragg peak occurs just beyond $k$ = 14$\pi$.  The constructed system in Figure \ref{fig:stealth3Dsplit} suppresses scattering for a range of $k$ surrounding 14$\pi$.  Because the intermediate set is free to fluctuate, the total number of constrained wave vectors in this stealth configuration is less than that of a crystalline configuration.  The stealth region away from the origin can be shifted to larger $k$ and can be tailored in magnitude.

The scattering of the intermediate set of wave vectors can also be controlled using Eq.\ (\ref{eq:phitilde}).  One such configuration that can be developed is a stealth/ideal gas hybrid.  In these many-particle systems, there is behavior typical of crystalline solids at selected regions of reciprocal space and ideal gas at others.  In Figure \ref{fig:stealth3DsplitID}, a 500 particle system in three dimensions is constrained so that $S(k)$ = 0 for  $0<|{\bf k}| \le 8.8\pi$ and $13\pi < |{\bf k}| \le 14.8\pi$. The intermediate set is constrained to $S(k)$ = 1.

The characteristics of the radial distribution function vary depending on the constraints of the intermediate set.  In all cases, the radial distribution function shows weak neighbor peaks before approaching unity. The second neighbor peak, though, is stronger than the first neighbor peak, a trait uncommon to most conventional many-body systems. The extent of the repelling-core region varies depending on the chosen value of $K_1$ and the extent to which the intermediate set is controlled, evidenced by Figures \ref{fig:stealth3Dsplit} and \ref{fig:stealth3DsplitID}.  Further reducing the the value to which the intermediate set is controlled is likely to increase the repelling-core region given the relation between $S(k)$ and $g_2(r)$ in Eq.\ (\ref{eq:sktogr}).

Spherically constrained stealth particle patterns consisting of 168 particles with the intermediate set respectively unconstrained and constrained are shown in Fig.\ \ref{fig:stealth2Dcompare}. To serve as a basis of comparison, a realization of a wavy crystalline configuration generated by suppressing scattering for all wave vectors up to 22$\pi$ is also shown.  The wavy crystalline material, Fig.\ \ref{fig:stealth2Dcompare}a, stands in sharp contrast to the stealth materials since particles tend to align in well-patterned strings.  The spherically constrained stealth materials lack any order. With the intermediate region uncontrolled, Fig.\ \ref{fig:stealth2Dcompare}b, particles tend to align into weak ``strings.''  Creating a stealth/ideal gas hybrid, Fig.\ \ref{fig:stealth2Dcompare}c, decreases the tendency to align in weak strings. The diameter of the circular window is equivalent to the length scale of $K_3$, indicating the ability to impose system features with a specified length scale. 

\section{Results for Super-Ideal Gases}
\label{sec:superideal}
 
Super-ideal gases were constructed at various $\chi$ values for $d=1,2$, and 3.  The maximum attainable $\chi$ value varied depending on spatial dimension and system size but was generally near $\chi = 0.95$ for most initial configurations.  At $\chi$ near unity, the minimization routine sometimes failed to find a global minimum of $\Phi$. However, in $d=3$ and $\chi$ near unity, the success rate for finding global minimum was much improved over that of lower dimensions.  

The results from all dimensions studied have similar characteristics.  For a single realization, the structure factor is exactly unity for $k<K$. Outside of the constrained region, the structure factor seemingly fluctuates about unity.  However, for an ensemble of super-ideal gases, the structure factor had a small peak immediately beyond $K$ that slowly decayed to unity.  This small peak is unexpected since we impose no constraints on $S(k)$ for wave vectors outside $K$ and would expect that correlations not exist at these $k$. The construction of super-ideal gases reveals a subtle coupling between $S(k)$ within the constrained region and $S(k)$ outside the constrained region that manifests upon ensemble averaging.  For $d=1$, $2$, and $3$, $S(k)$ never exceeds a value of 1.25, 1.18, and 1.10 respectively when ensemble averaged.  For $\chi<0.4$ and $\chi>0.96$, the peak generally decays to unity rapidly.   For $0.4 \leq \chi \leq 0.96$, the peak is rather long ranged, decaying much more slowly.  Figure \ref{fig:super3DhighANDlow} shows the ensemble-averaged $g_2(r)$ and $S(k)$ for two systems containing 500 particles in $d=3$.

The radial distribution function has characteristics common across dimensions studied. For $r>0.1L_x$, $g_2(r)$ shows very small fluctuations about unity. The local behavior of $g_2(r)$, $r< 0.1L_x$, varies depending on $\chi$ and is most sensitive to $\chi$ in $d=3$.  Figure \ref{fig:super3DhighANDlow} shows that super-ideal gases at $\chi=0.90667$ exhibit severe local clustering as the radial distribution function has a contact value $g_2(0)$ near 7.  However, at $\chi=0.98967$, the structure more closely resembles an ideal gas.  Figure \ref{fig:superGRpeak} tracks the contact value of $g_2(r)$ for various $\chi$.  At significantly large $\chi$, local clustering is suppressed and the super-ideal gas structure closely resembles that of an ensemble of ideal gas configurations. An interesting consequence is that the local structure of a super-ideal gas at very small $\chi$ resembles that of a super-ideal gas at very large $\chi$.

Two-dimensional particle patterns reveal subtle differences between super-ideal gas configurations and Poisson distributions.  Figure \ref{fig:super2Dconfig} compares a realization of Poisson distribution of 418 particles and a super-ideal gas at $\chi$=0.90.  At certain $\chi$, super-ideal gases exhibit local order and a tendency to align into weak ``strings'' that is best revealed only upon ensemble averaging.  At smaller $\chi$, super-ideal gases are particularly difficult to discern from a Poisson distribution.

\section{Results for Equi-Luminous Materials}
\label{sec:lumin}
Equi-luminous materials represent a broad class of materials that scatter light equally intensely for a set of wave vectors, where a super-ideal gas is a special case.  We choose to focus here on equi-luminous materials whose scattering in the small-$k$ region is much more intense relative to that of an ideal gas.  For all cases considered, we observe qualitative similarities in the ensemble-averaged radial distribution function and structure factor.  The structure factor is exactly equal to the chosen constant for all $k<K$. Beyond $K$, the ensemble-averaged structure factor decays to unity very slowly at a rate dependent on the specified constant in constrained region. Figure \ref{fig:lumin2DSK} compares $S(k)$ for several equi-luminous configurations containing 168 particles for $d=2$.  These configurations have constrained $S(k)$ values that are exactly 1,2,3, and 4, respectively, for $\chi=0.34523$.  Achieving $\chi$ above 0.37 for $S(k)$ = 4.0 for $d=2$ proved challenging as the minimization procedure often failed to find global minima.

In real space, we observe that constraining $S(k)$ to be increasingly large has no affect on the long-range behavior of $g_2(r)$.  Generally, for $r>0.1L_x$, $g_2(r)$ averages near unity.  However, for $r<0.1L_x$, strong local correlations rapidly vanish for increasing $r$.  By increasing the constrained value of $S(k)$, the contact value of $g_2(r)$ increases significantly but does not change the large $r$ behavior, which remains near unity. Figure \ref{fig:lumin2DGR} demonstrates the behavior of $g_2(r)$ corresponding to the systems described above.

Realizations of equi-luminous materials demonstration the aggregation of particles common to this class of equi-luminous materials. Figure \ref{fig:lumin2Dconfig} shows configurations for which $S(k)$ = 2 and $S(k)$ = 4 for $\chi$ = 0.37.  Particles tend to cluster and align into well-formed strings, or filamentary structures, as opposed to clustering radially.  Filamentary structures arise in astrophysical systems, particularly for the distributions of galaxies. \cite{macgillivray1986fso}  For a larger target $S(k)$, the aggregation is increasingly severe and particles nearly stack on top of each other.

In the extreme case of targeting $S(k)$ toward its maximum value of $N$ for very small $\chi$, particles tend to collapse upon each other, yielding an overall rescaling of the system length scale. It should be noted that in these extreme cases, global minima were rarely found and required successive iterations to achieve optimality. For example, in seeking a ground-state configuration with $S(k)=10$ for $\chi=0.01785$, the ground-state configuration in which $S(k)=8$ for $\chi=0.01785$ was used as the initial condition.

\section{Conclusion and Discussions}
\label{sec:conclusions}
Previous studies concerning constraints on the collective coordinates of particle configurations \cite{uche2006ccc, uche2004ccd, fan1991ccd} were suggestive that potentials defined by Eqs.\ (\ref{eq:phi}) and (\ref{eq:phitilde}) yield classical disordered ground-state configurations. These studies restricted the system size to small periodic boxes consisting of at most 12 particles in $d=1$,\cite{fan1991ccd} 418 particles in $d=2$, \cite{uche2004ccd} and 500 particles in $d=3$.\cite{uche2006ccc}  In this investigation, we find that increasing the system size while fixing $\rho$ and $\chi$ does not affect the structure of resulting configurations, suggesting that there are no long-range correlations in the infinite-volume limit.  Constructing ground state materials via MINOP for systems significantly larger than that presented in Section \ref{sec:stealth} becomes computationally challenging and currently is the limitation on system size.  Regardless, our numerical evidence for disordered ground states in the infinite-volume limit corroborates recent analytical work that suggests the existence of energetically degenerate and aperiodic ground states in the infinite-volume limit via a similar potential.\cite{suto2005cgs, torquato2008ndr} 

Three novel classes of ground-state materials with potential radiation scattering applications have been introduced: stealth materials, super-ideal gases, and equi-luminous materials. Each provides an unique opportunity to impose an underlying structure for a ground-state configuration with a known potential. With stealth materials, we have the precise control to suppress scattering at specified wavelengths. For a single square-mound $V({\bf k})$, increasing the number of constraints on wave vectors near the origin drives systems toward crystallinity. \cite{fan1991ccd, uche2004ccd, uche2006ccc}  However, by introducing $V({\bf k})$ with two square mounds and choosing to suppress scattering in two disconnected regions of reciprocal space, we disrupt the drive toward crystallinity.  Particle patterns then have strong local correlations with an imposed length scale, but lack long-range order.  It is interesting to note that all stealth ground-state configurations generated in this investigation are also hyperuniform point patterns; i.e., configurations that suppress density fluctuations in the infinite-wavelength limit.  In choosing which regions of reciprocal space to be stealthy, we exhibit the unique ability to control and suppress density fluctuations at specified wavelengths. 

Equi-luminous materials investigated in Section \ref{sec:lumin} demonstrate ground-state structures that scatter light for a set of wavelengths more strongly than an ideal gas.  In choosing the the number of constrained wave vectors and the value of $S(k)$, one can design ground-state configurations that have increasingly clustered behavior for increasing $\chi$ and $S(k)$. By adjusting $K$, we can impose correlations over a certain length scale.  

The differences between super-ideal gases and ideal gases are not intuitively obvious. The resulting local structure deviates significantly from that of an ideal gas as there is some degree of local clustering, with the exception of $\chi$ near 0 and $\chi$ near 1.  The propensity for super-ideal gases to exhibit local clustering suggests that, among the many energetically degenerate global minima in the landscape, configurations with some degree of clustering dominate relative to those that do not.  We also find that this local order is particularly sensitive to the number of constraints imposed suggesting that the energy landscape changes significantly for small changes in $\chi$. Despite the local clustering, the resulting configurations exhibit no long-range order.

One potential application of this procedure is in addressing the realizability question. Despite receiving significant attention, only necessary conditions have been placed on the pair correlation function $g_2(r)$ and its corresponding structure factor $S(k)$ for the realizability of a point process.\cite{torquato2002rhm, torquato2002csr, costin2004cpd,  stillinger2004pcf, stillinger2005rii, torquato2006ncl} General sufficient conditions have yet to be developed.  This method can potentially address realizability from reciprocal space, which stands in contrast to real-space numerical reconstruction techniques.\cite{yeong1998rrm, cule1999grm, sheehan2001gms, crawford2003acf, uche2006rpc} A limitation in its use for the question of realizability is the peaking phenomenon in $S(k)$. In all cases studied, $S(k)$ has a peak immediately beyond $K$, suggesting collective coordinates are not necessarily independent of each other and that the nonlinearity of the coordinate transformation plays an important role in constraining collective coordinates.  Additionally, our results show that the magnitude of the peak is reduced for higher dimensions, suggesting that dimensionality plays a key role in the coupling among collective coordinates. The understanding of constraints on collective coordinates is incomplete. New questions arise: how are the wave vectors beyond $K$ influenced by constraints below $K$ and what role does dimensionality play in the peaking phenomenon?   

Another possible application of this procedure is to produce ground-state configurations that are ordered or quasiperiodic over a specified length scale for the design of photonic materials.  Materials with a photonic band gap are of significant interest due to their technological applications, \cite{john1987slp, yablonovitch1987ise, yablonovitch1993pbg} and materials with desired band gaps have been designed.\cite{zhang2000rpb, asatryan1999edw} From first-order perturbation theory, scattering of radiation is related to the photonic band gap of a material.\cite{johnson2002ptm} By targeting a structure factor that is maximal for certain wave vectors and zero for others, we may be able to construct ground-state structures with a specified desired photonic band gap to a first-order approximation.  

In the present investigation, the potential $V({\bf k})$ was chosen to be a square mound with compact support at $K$.  However, the constructed ground-state configurations are equivalent to the ground-state configurations associated with a broad class of $V({\bf k})$.  If a ground-state configuration of density $\rho$ is constructed by minimizing Eq.\ (\ref{eq:phi}) or Eq.\ (\ref{eq:phitilde}) to its global minimum value, Eq.\ (\ref{eq:min}) or zero respectively, then the ground-state configuration is also a ground state at density $\rho$ for any positive $V({\bf k})$ that is bounded with the same compact support at $K$.

Recently derived duality relations allow us to identify further degeneracies among ground states constructed via Eq.\ (\ref{eq:phi}) and provide bounds on the minimum energy for various forms of $v({\bf r})$.  These duality relations link the energy of configurations for a bounded and integrable real-space pair potential function to that of its Fourier transform and provide a fundamental connection between the ground states of short-range pair potentials to those with long-range pair potentials.  More specifically, if a Bravais lattice structure is the ground state energy $U_{min}$ for a pair potential function $v({\bf r})$,  then $U_{min}$ provides an upper bound on the ground state energy $\tilde{U}_{min}$ of a system interacting via the Fourier-transform of the pair potential function $V({\bf k})$ at the associated reciprocal density.\cite{torquato2008ndr}  In using Eq.\ (\ref{eq:phi}), we used a long-ranged, damped, and oscillatory $v({\bf r})$, whose Fourier transform is a square mound $V({\bf k})$.  For $\chi \leq 1$ and at a given density $\rho$, this $v({\bf r})$ had a Bravais lattice as a ground-state structure, and for sufficiently large $\chi$, a Bravais lattice becomes its unique ground-state structure at a certain density.   By these duality relations, we know that the minimum energy of this Bravais lattice system, Eq.\ (\ref{eq:min}), is the upper bound on the ground-state energy of systems at density $\rho^*=\rho^{-1}(2\pi)^{-d}$ interacting via $V({\bf k})$.\cite{torquato2008ndr}

Targeting specific forms of the structure factor with Eq.\ (\ref{eq:phitilde}) requires up to four-body interactions in real space to achieve these as ground states.  However, it would be particularly useful to construct disordered ground states using short-range, two-body potentials.  Developing effective short-range, pair interactions, potentially via the Ornstein-Zernike formalism,\cite{hansen2006tsl} to achieve desired scattering properties remains a potential future direction.

\begin{acknowledgments}
The authors thank Obioma Uche and Chase Zachary for for helpful discussions regarding several aspects of this work. S. T. thanks
the Institute for Advanced Study for their hospitality during
his stay there. This work was supported by the Office of Basic Energy Sciences, U.S. Department of Energy, under Grant DE-FG02-04-ER46108.
\end{acknowledgments}

\appendix

\section{Energy Landscape Analysis}
\label{sec:energy}
For large $\chi$, the procedure sometimes terminates at a potential higher than the absolute minimum, failing to meet our criterion for global minimum.  Generally this has been attributed to local minima in the energy landscape associated with Eqs.\ (\ref{eq:phi}) and (\ref{eq:phitilde}) and the performance of the minimization algorithm. \cite{uche2006ccc}  Since little is known about these landscapes, we shed some light unto them and justify the discrepancy between the performance of the various algorithms.  

\subsection{Energy Landscape}
The landscape of $\Phi$ can be visualized for a system of three particles on a unit line.  Imposing the constraint so that $S({\bf k})|_{{\bf k}=\pm2\pi}= \frac{2}{N}D$, the energy, $\Phi$, becomes 
\begin{equation}
\Phi = \{ \cos[2\pi(x_1-x_2)] + \cos[2\pi(x_1-x_3)] + \cos[2\pi(x_2-x_3)] + \frac{3}{2} - D\}^2.
\label{eq:threepart}
\end{equation}
$D$ must be within the range of realizability for $S(k)$ (i.e., $0 \le D \le \frac{N^2}{2}$). The energy landscape possesses translational freedom, so we can fix $x_1$ at the origin, $x_1$=0.  Plotting Eq.\ (\ref{eq:threepart}) versus particle coordinates for the case of $D$=0 and $D$=1 provides a simple picture of the energy landscape.  Figures \ref{fig:lszero} and \ref{fig:lssuper} show the energy landscape and contour plot for $D$=0 and $D$=1 respectively.

For a stealth configuration, $D=0$, Figure \ref{fig:lszero}, there is exactly one solution, corresponding to the crystalline arrangement of particles, that is a global minimum of $\Phi$. The global maximum is a stacking of all particles onto the origin. For $D>0$, Figure \ref{fig:lssuper}, there is a family of degenerate ground-state solutions located in a ring around configurational points corresponding to the periodic arrangement of particles.  For this system, a single ground state exists only for $D=0$ or $D=\frac{N^2}{2}$.  For all other $D$, there exist degenerate ground states.  Taylor expansion about the minima for when $D$=0 indicates the landscape is quartic in nature. When $D \neq 0$, local maxima are located at configurational points associated with the periodic arrangement of particles, as evident in Figure \ref{fig:lssuper}.  As $D$ is increased, the local maxima are found increasingly higher in the landscape.  Above $D$=1.5, the periodic arrangements of particle become global maxima and the origin becomes a local maximum. 

Visualization confirms that the energy landscape is nontrivial, particularly for $D \neq 0$.  Local minima were not found despite our experience with the minimization procedure.  Local minima can be found by constraining the next wave vector. For example, constructing a system for $D=1$ at $\chi$=0.667 for a three-particle system introduces local minima.  If $D\neq0$ and $D\neq \frac{N^2}{2}$, $\Phi$ cannot be minimized to zero at $\chi$=0.667.  The crystalline arrangement achieves $\Phi$=0 for $D=0$ and the stacking of particles achieves $\Phi$=0 for $D=\frac{N^2}{2}$.

\subsection{Stationary Point Characterization and Capture Basins}
To enumerate and characterize the stationary points (local/global minimum or maximum, or saddle point) of the energy landscape $\Phi$, Eq.\ (\ref{eq:phi}), we introduce a nonnegative function, $f({\bf x})$, where ${\bf x}$ refers to particle coordinates:
\begin{equation}
\label{eq:phidotphi}
f({\bf x}) = \nabla\Phi \cdot \nabla\Phi.
\end{equation}
The global minima of $f({\bf x})$ have a value of zero and correspond to stationary points of the energy landscape $\Phi$.  Capture basins are the regions of phase space that, upon minimizing $f({\bf x})$, yield each stationary point.  For three particles on a line, the stationary points are countable from the visualization of the landscape, and this provides a basis for comparison.  We use steepest descent since it will generate nonoverlapping capture basins and we compare against MINOP.  If at a certain point in the energy landscape MINOP's trust region overlaps with a different capture basin, it may favor a search direction that enters the overlapping capture basin and lead to a different stationary point.  

Table \ref{tab:threepart} summarizes the results of the capture basin analysis for a three particle one-dimensional system constrained at $k$=1 and $k$=2 from several hundred trials. Random positions are assigned and $f({\bf x})$ is minimized, yielding a stationary point.  Many times $f({\bf x})$ was not minimized to a global minimum indicating that a stationary point of $\Phi$ was not found. In the table, the number of capture basins refers to the countable number of basins in the landscape with one basin associated with each stationary point, without accounting for particle permutations.  The fractions in the chart represent the fraction of the total number of stationary points found corresponding to each stationary point, or alternatively, the fraction of phase space that corresponds to a specific capture basin for each stationary point.

The capture basins for a two-dimensional system of 12 particles were found for all available values of $\chi$, for $D=0$. Because of the increased speed of MINOP, we only employ this algorithm to minimize $f({\bf x})$.  Upon generating a stationary point, the potential $\Phi$ was calculated and the eigenvalues of its associated Hessian matrix were found.  Global minima of $\Phi$ have nonnegative eigenvalues and $\Phi=0$.  Saddle points have at least one negative eigenvalue and $\Phi>0$. Interestingly no local minima, local maxima, or global maxima were found.  Table \ref{tab:12part} summarizes the capture basins and stationary points for 12 particles in two dimensions.  Trials that did not yield stationary points are excluded from the reported fractions.

No local maxima, global maxima, or local minima were found, however, they must exist in systems of sufficiently high dimension.  The most unstable saddles generally have a high potential exceeding the global minimum by $>10^2$, while the most stable saddles have potentials $\Phi$ that differed from a global minimum by O($10^{-1}$).  Our experience with finding local minima is consistent with these results.  When MINOP terminates at $\Phi$ above its global minimum value, which usually occurs when $N \gg 12$, the potential of the final configuration is about $10^{-8}$ to $10^{-2}$ above the global minimum value.  The strong correlation between the number of negative eigenvalues and potential is consistent with our experience with these algorithms.

The number of saddle points grows rapidly with the total dimensionality of the system (number of particles times spatial dimension), and grows more rapidly than any other stationary points.  In the three-particle system at $\chi$=0.667, we observe that local maxima are located near saddle points in the energy landscape. It is likely that minimizing $f({\bf x})$ introduces a bias toward finding saddle points of $\Phi$ rather than local maxima of $\Phi$. Because local maxima are outnumbered by saddle points and the search procedure may favor directions toward saddle points of $\Phi$, it likely that minimizing Eq.\ (\ref{eq:phidotphi}) would rarely generate local maxima for systems of high dimensionality. Global maxima are attained only by stacking particles in a single location. It is expected that this stationary point would be so greatly outnumbered by all other stationary points that minimizing $f(x)$ may never generate a global maximum.

These simple studies allow us to better justify the inferences regarding the landscape associated with $\Phi$.  Overconstraining the three particle system for targeting a super-ideal gas introduced local minima.  Increases to the system's dimensionality and to $\chi$ increased the number of saddle points and is expected to increase the number of local minima. Comparisons between MINOP and line search techniques indicate that the trust region often overlaps other capture regions which find global minima of $\Phi$ more effectively.

\newpage
\begin{table}
\caption{\label{tab:threepart}Fraction of phase space corresponding to capture basins of landscape $\Phi$ for three particles on a line as found by steepest descent (SD) and MINOP algorithms.}
\begin{ruledtabular}
\begin{tabular}{c|ccc|ccc}
&\multicolumn{3}{c|}{$\chi$=0.333} &\multicolumn{3}{c}{$\chi$=0.667} \\
\hline
          & \# of capture basins & SD     & MINOP & \# of capture basins & SD    & MINOP \\
\hline
Global Min&                     2& 0.414  &  0.290&                    3 & 0.234 &  0.146 \\
Saddle    &                     3& 0.440  &  0.530&                    6 & 0.420 &  0.376 \\
Global Max&                     1& 0.108  &  0.180&                    1 & 0.016 &  0.108 \\
Local Max &                      &        &       &                    2 & 0.048 &  0.014 \\
Failed to Converge&             -& 0.038  &  0.000&                    - & 0.282 &  0.356 \\ 
\end{tabular}
\end{ruledtabular}
\end{table}

\begin{table}
\caption{\label{tab:12part}Fraction of phase space corresponding to capture basins of landscape $\Phi$ for 12 particles in a unit square as found by the MINOP algorithm. }
\begin{ruledtabular}
\begin{tabular}{cccccc}
\multicolumn{3}{c}{} &\multicolumn{2}{c}{Fraction of Stat.\ Pts.\ } \\
$\chi$ & \# Trials & \# Stat Points  &Global Min  &Saddles \\ 
\hline
0.083  & 599  & 599 & 0.8881 & 0.1135 \\
0.167  & 600  & 596 & 0.3591 & 0.6049 \\
0.250  & 700  & 558 & 0.0663 & 0.9337 \\
0.417  & 697  & 431 & 0.0000 & 1.0000 \\
0.500  & 795  & 372 & 0.0000 & 1.0000 \\ 
0.583  & 899  & 433 & 0.0000 & 1.0000 \\
\end{tabular}
\end{ruledtabular}
\end{table}

\newpage
\clearpage

\begin{figure}
\centering
\caption{\label{fig:stealth3DinfSK} Structure factor for stealth ground states for $d$=3, $\rho$=108, and $\chi$=0.05.  Increasing the system size from $N$=108 to $N$=6912 does not affect the scattering characteristics. The potential energy was minimized to within 10$^{-17}$ of its absolute minimum.}
\end{figure}

\begin{figure}
\centering
\caption{\label{fig:stealth3DinfGR} Radial distribution function for stealth ground states for $d$=3, $\rho$=108, and $\chi$=0.05.  Increasing the system size from $N$=108 to $N$=6912 does not affect the resulting local structure. The potential energy was minimized to within 10$^{-17}$ of its absolute minimum.}
\end{figure}

\begin{figure}
\centering
\caption{Ensemble-averaged $S(k)$ for stealth ground states consisting of 500 particles in a unit cube. At $\chi >$ 0.45, $S(k)$ begins to oscillate while damping to unity.  (a) $\chi$ = 0.11333, 250 realizations, (b) $\chi$ = 0.25000, 50 realizations, (c) $\chi$ = 0.45467, 50 realizations, (d) $\chi$ = 0.54867, 4 realizations. The potential energy was minimized to within 10$^{-17}$ of its global minimum.}
\label{fig:stealth3Dsk}
\end{figure}
  
\begin{figure}
\centering
\caption{Ensemble-averaged $g_2(r)$ for stealth ground states of 500 particles in a unit cube. At $\chi >$ 0.45, $g_2(r)$ oscillates about unity with a shorter wavelength than observed for smaller $\chi$ values. (a) $\chi$ = 0.11333, 250 realizations, (b) $\chi$ = 0.25000, 50 realizations, (c) $\chi$ = 0.45467, 50 realizations, (d) $\chi$ = 0.54867, 4 realizations. The potential energy was minimized to within 10$^{-17}$ of its global minimum.}
\label{fig:stealth3Dgr}
\end{figure}

\begin{figure}
\centering
\caption{Stealth particle patterns of 168 particles in two dimensions. (a) $\chi$ = 0.04167, (b) $\chi$ = 0.20238.  The bar below each graph and the circular window represent the characteristic length associated with $K$. Both systems are disordered but at higher $\chi$, particles tend to spread away from each other. The potential energy was minimized to within 10$^{-17}$ of its global minimum.}
\label{fig:config2dstealth}
\end{figure}

\begin{figure}
\caption{(a) Ensemble-averaged $g_2(r)$ and (b) $S(k)$ for a stealth material of 500 particles in a unit cube. 25 realizations. $S(k)$ = 0 for  $0<|{\bf k}|\le8.8\pi$ and $13\pi < |{\bf k}| \le 14.8\pi$ and with the intermediate set unconstrained.  The potential energy was minimized to within 10$^{-17}$ of its global minimum.}
\label{fig:stealth3Dsplit}
\end{figure}

\begin{figure}
\caption{(a) Ensemble-averaged $g_2(r)$ and (b) $S(k)$ for a stealth material of 500 particles in a unit cube.  $S(k)$ = 0 for  0 $< |{\bf k}| \le 8.8\pi$ and $13\pi < |{\bf k}| \le 14.8\pi$ and with the intermediate set constrained to $S(k)$ = 1. 10 realizations. The potential energy was minimized to within 10$^{-17}$ of its global minimum.}
\label{fig:stealth3DsplitID}
\end{figure}

\begin{figure}
\caption{(a) Wavy crystalline configuration generated by constraining $S(k)$ = 0 for all ${\bf k} \le 22\pi$.  (b) 
Stealth material generated by constraining $S(k)$ = 0 for $0< |{\bf k}| \le 10\pi$ and $20\pi < |{\bf k}| \le 26\pi$ and the intermediate set is unconstrained. (c) Stealth material generated by constraining $S(k)=0$ for $0 < |{\bf k}| \le 10\pi$ and $18\pi < |{\bf k}| \le 22\pi$ and with the intermediate set constrained to $S(k) = 1$. The line beneath the figure and the circle in the figure approximately represent the length scale of $K_3$. The potential energy was minimized to within 10$^{-17}$ of its global minimum. }
\label{fig:stealth2Dcompare}
\end{figure}

\begin{figure}
\centering
\caption{Ensemble-averaged $g_2(r)$ and $S(k)$ for a super-ideal gases in three dimensions. 500 particles in unit cube.  $\chi$ = 0.90667, 30 realizations, and $\chi$ = 0.98933, 12 realizations.  The dashed line shows location of unity for each structure factor. The potential energy was minimized to within 10$^{-17}$ of its global minimum.}
\label{fig:super3DhighANDlow}
\end{figure}

\begin{figure}
\centering
\caption{The value of radial distribution function in the first bin for 500 particles in a unit cube. The contact value $g_2(0)$ increases initially for small $\chi$. However, for large $\chi$, local clustering is suppressed.  Each data point represents at least 30 realizations, except for $\chi=0.98967$ which represents 12 realizations.}
\label{fig:superGRpeak}
\end{figure}

\begin{figure}
\centering
\caption{Particle configurations of 418 particles. (a) Poisson point process, (b) super-ideal gas, $S(k)$ = 1, $\chi$ = 0.90. Ensembles of super-ideal gases reveal the presence of local clustering.}
\label{fig:super2Dconfig}
\end{figure}

\begin{figure}
\caption{Ensemble-averaged $S(k)$ for equi-luminous ground states consisting of 168 particles for $d=2$. $\chi$ = 0.34523, 50 realizations. The potential energy was minimized to within 10$^{-17}$ of its global minimum.}
\label{fig:lumin2DSK}
\end{figure}

\begin{figure}
\caption{Ensemble-averaged $g_2(r)$ for super ideal gas and equi-luminous ground states consisting of 168 particles, $d$=2, $\chi$ = 0.34523, 50 realizations.  Clustering near the origin increases for increasing the constrained $S(k)$ value, which are 1, 2, 3, and 4 respectively. The potential energy was minimized to within 10$^{-17}$ of its global minimum.}
\label{fig:lumin2DGR}
\end{figure}

\begin{figure}
\centering
\caption{Ground-state configurations of 168 particles (a) $S(k)$ = 2, $\chi$ = 0.34523, and (b) $S(k)$ = 4, $\chi$ = 0.34523.  Particle clustering increases when constrained to higher $S(k)$.}
\label{fig:lumin2Dconfig}
\end{figure}

\begin{figure}
\caption{Energy landscape and contour plot associated with minimizing Eq.\ (\ref{eq:threepart}) for the first wave vector. Three particles on a unit line, $D=0$, $\chi=0.333$.  Ground state configurations are to the periodic arrangement.}
\label{fig:lszero}
\end{figure}

\begin{figure}
\caption{Energy landscape and contour plot associated with minimizing Eq.\ (\ref{eq:threepart}) for the first wave vector. Three particles on a unit line, $D=1$, $\chi=0.333$. Ground state configurations are a set of configurations at the $\Phi$ = 0 ring around the integer lattice points.}
\label{fig:lssuper}
\end{figure}

\clearpage
\newpage

\setcounter{figure}{0}

\begin{figure}
\includegraphics[width = \textwidth, clip=true]{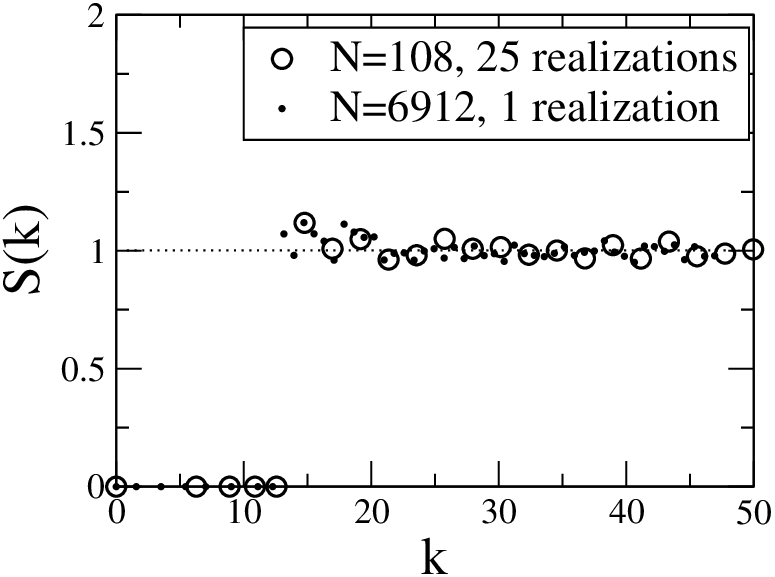}
\centering
\caption{Batten, Stillinger, Torquato}
\end{figure}

\clearpage
\newpage

\begin{figure}
\includegraphics[width = \textwidth, clip=true]{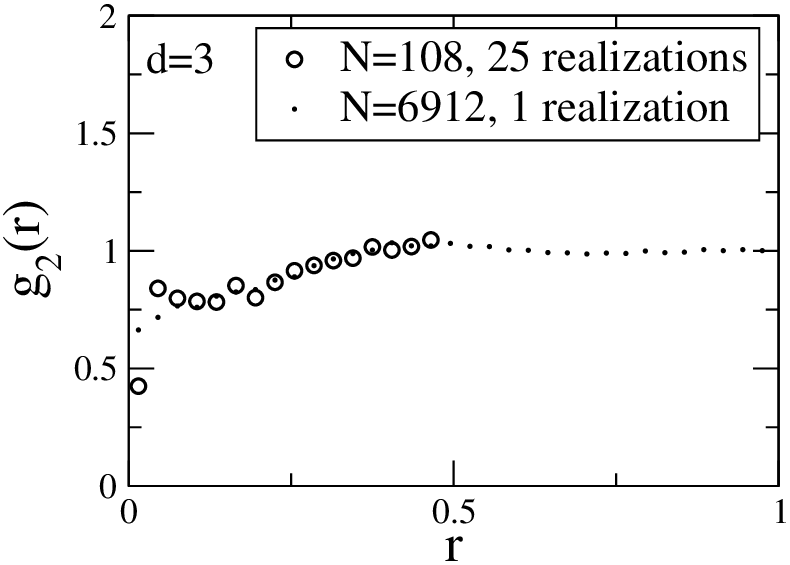}
\centering
\caption{Batten, Stillinger, Torquato}
\end{figure}

\clearpage
\newpage

\begin{figure}
\includegraphics[width=\textwidth, clip=true]{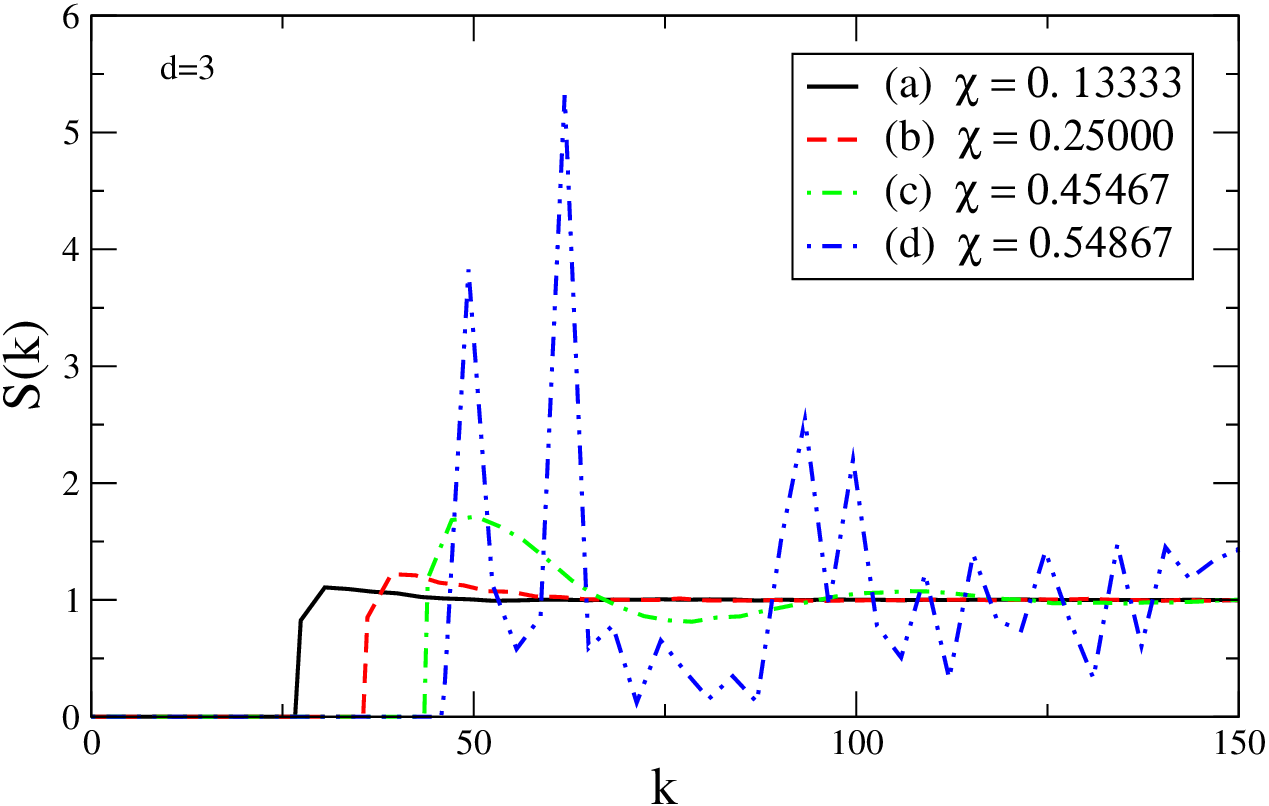}
\centering
\caption{Batten, Stillinger, Torquato}
\end{figure}

\clearpage
\newpage
  
\begin{figure}
\includegraphics[width=\textwidth,clip=true]{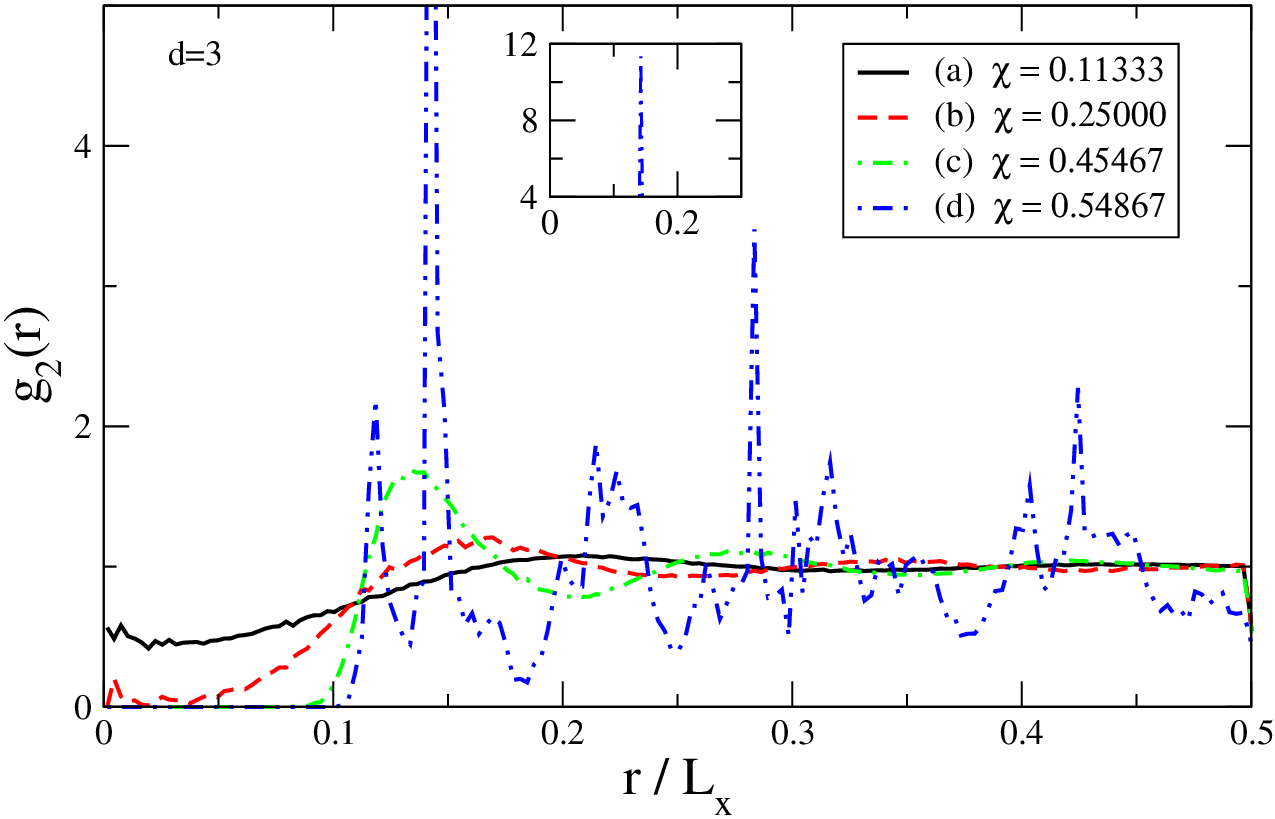}
\centering
\caption{Batten, Stillinger, Torquato}
\end{figure}

\clearpage
\newpage

\begin{figure}
\includegraphics[width=\textwidth,clip=true]{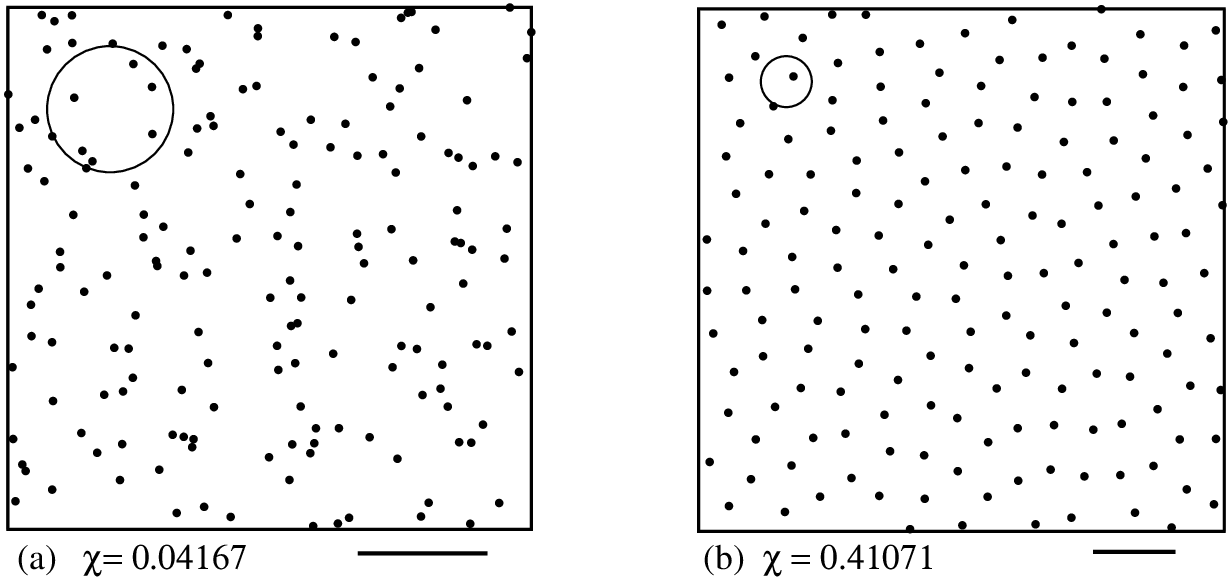}
\centering
\caption{Batten, Stillinger, Torquato}
\end{figure}

\clearpage
\newpage

\begin{figure}
\includegraphics[width=\textwidth, clip=true]{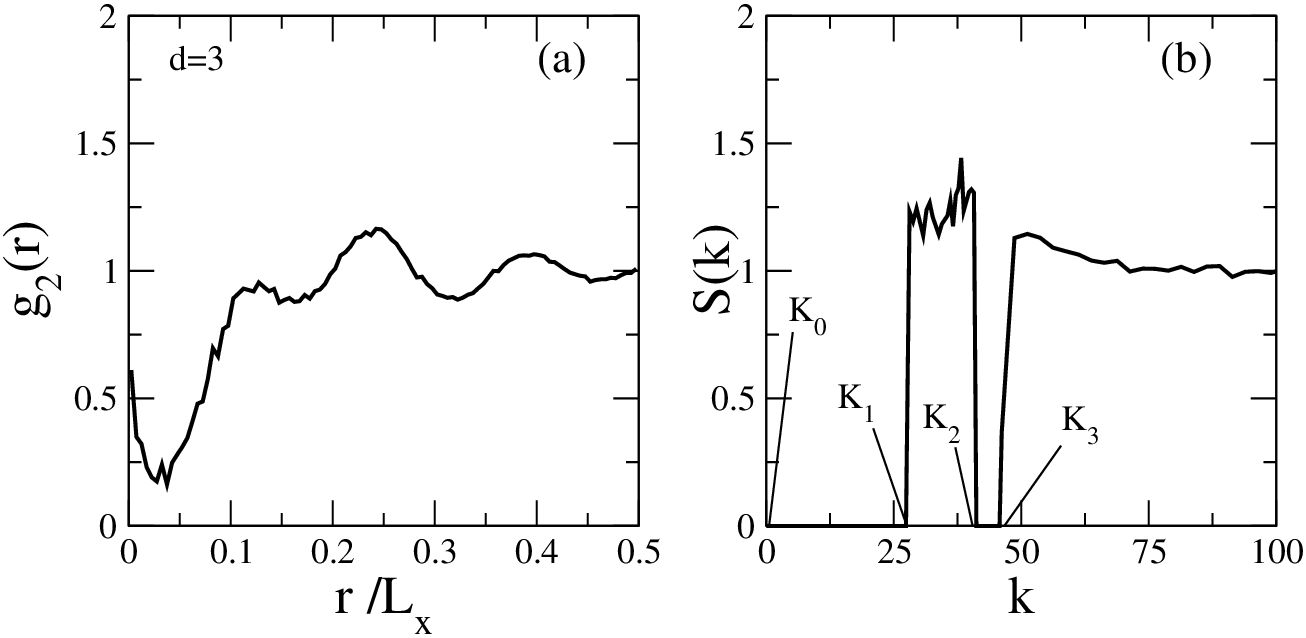}
\centering
\caption{Batten, Stillinger, Torquato}
\end{figure}

\clearpage
\newpage

\begin{figure}
\includegraphics[width=\textwidth, clip=true]{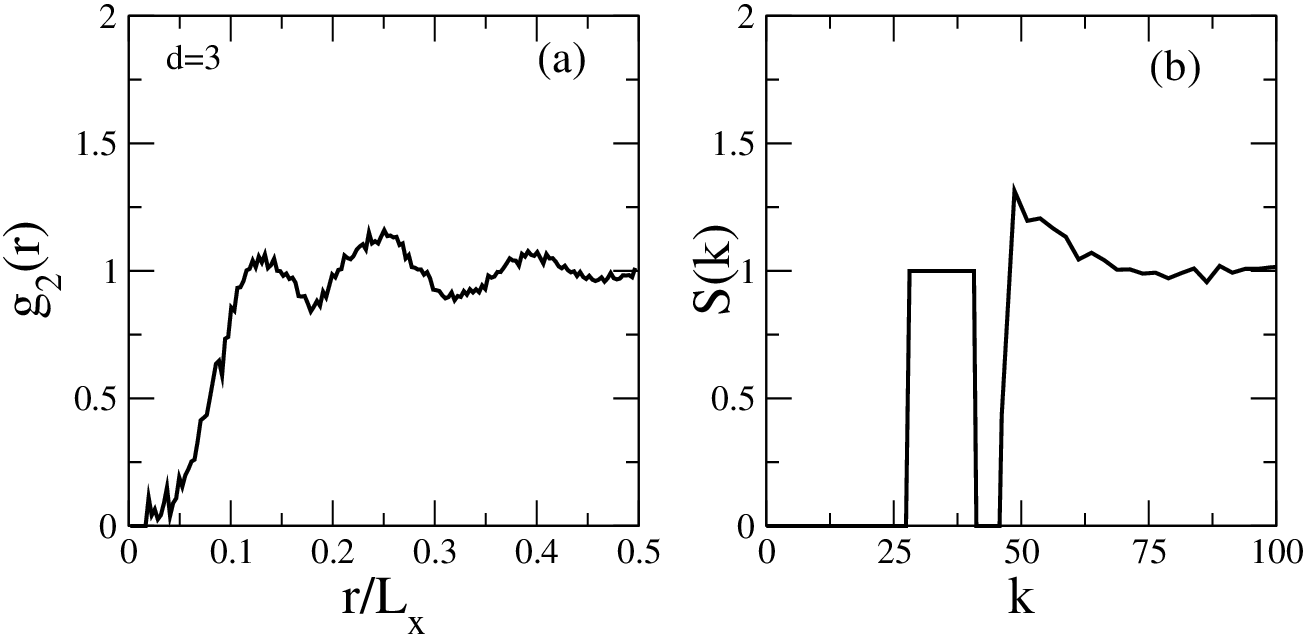}
\caption{Batten, Stillinger, Torquato}
\end{figure}

\clearpage
\newpage

\begin{figure}
\includegraphics[width=\textwidth, clip=true]{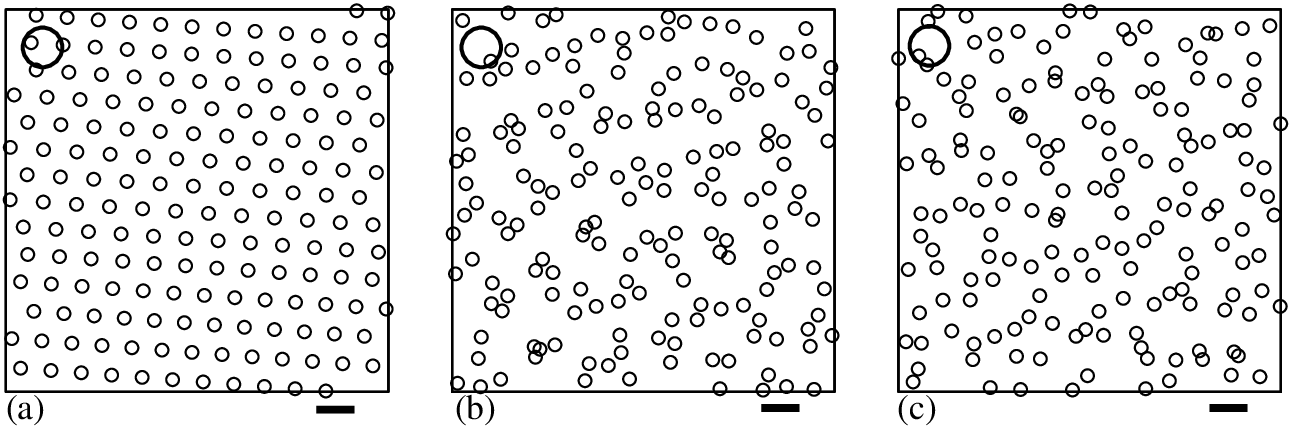}
\caption{Batten, Stillinger, Torquato}
\end{figure}

\clearpage
\newpage

\begin{figure}
\includegraphics[width=\textwidth, clip=true]{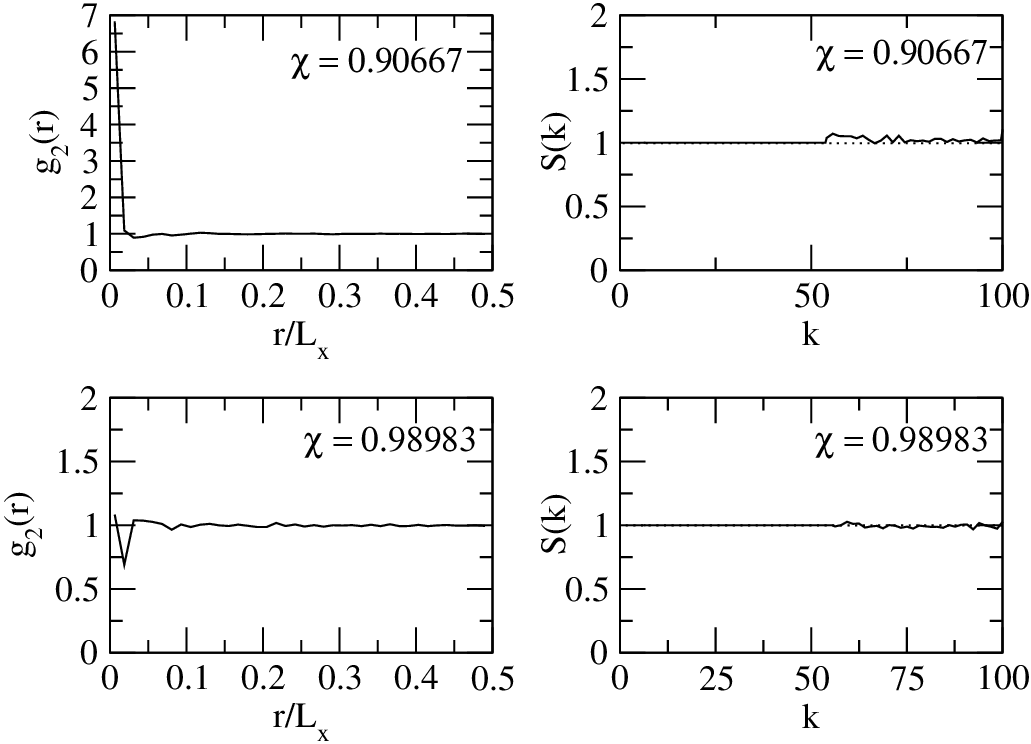}
\centering
\caption{Batten, Stillinger, Torquato}
\end{figure}

\clearpage
\newpage

\begin{figure}
\includegraphics[width=\textwidth,clip=true]{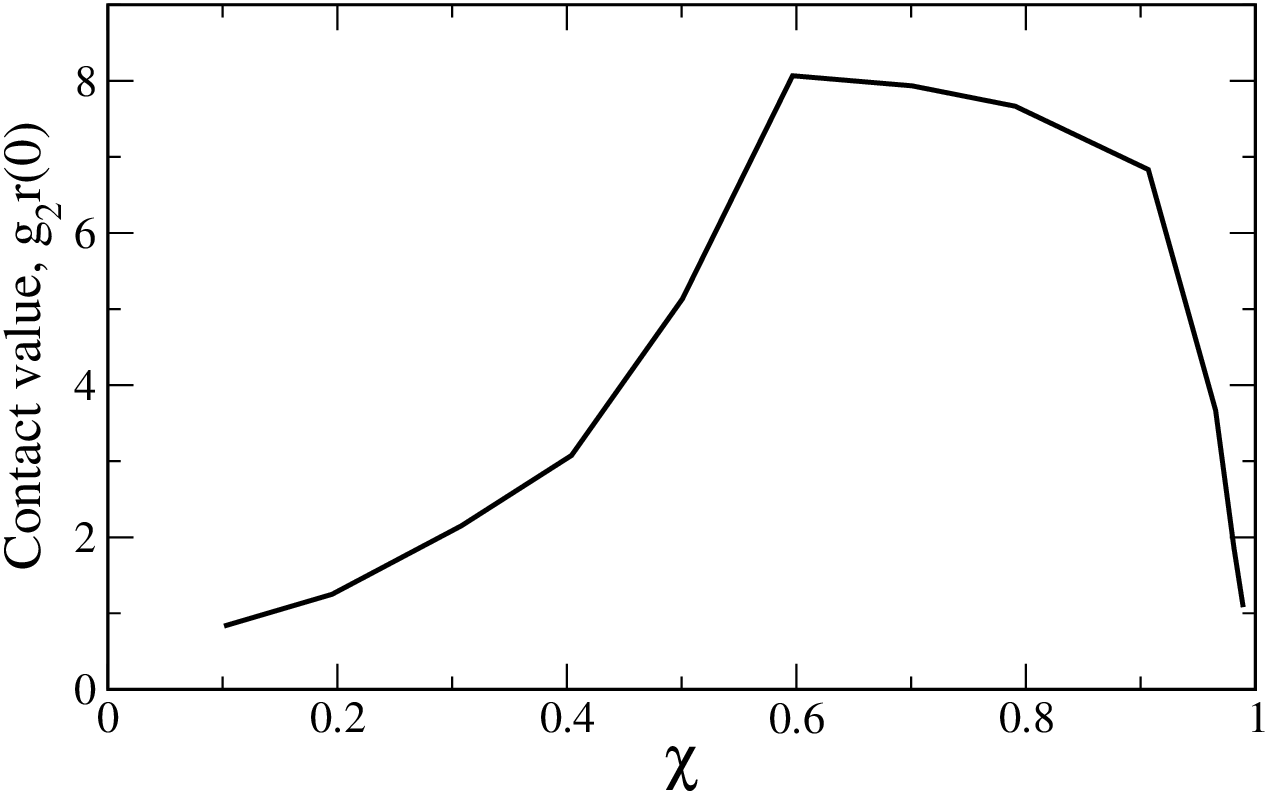}
\centering
\caption{Batten, Stillinger, Torquato}
\end{figure}

\clearpage
\newpage

\begin{figure}
\includegraphics[width=\textwidth,clip=true]{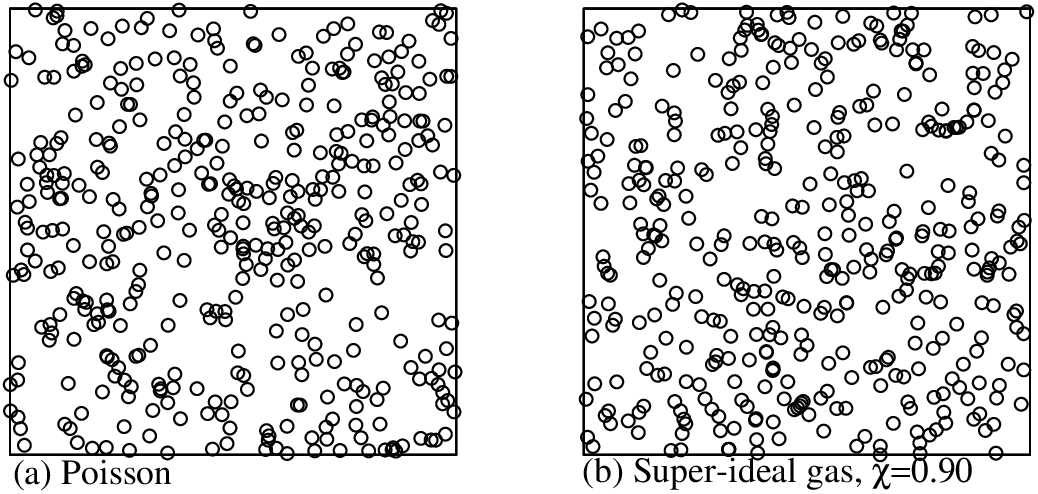}
\centering
\caption{Batten, Stillinger, Torquato}
\end{figure}

\clearpage
\newpage

\begin{figure}
\includegraphics[width=\textwidth,clip=true]{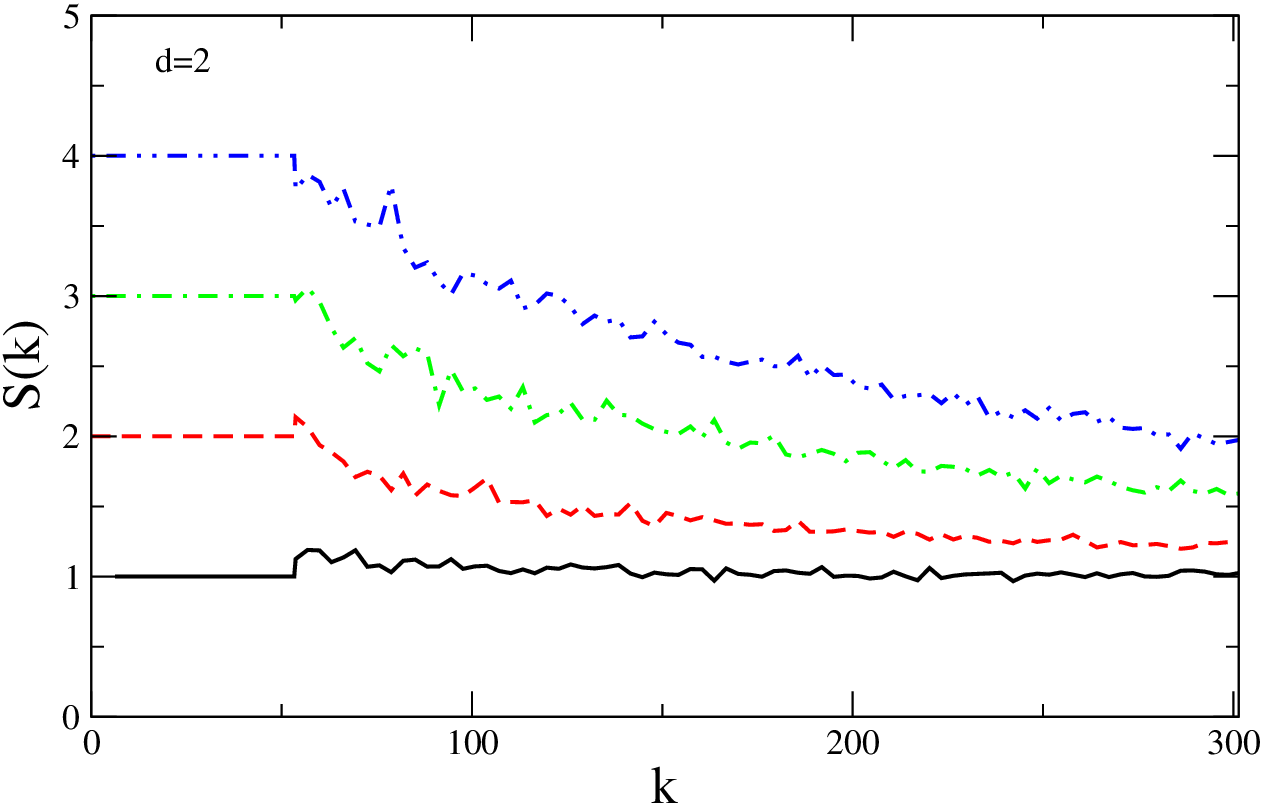}
\caption{Batten, Stillinger, Torquato}
\end{figure}

\clearpage
\newpage

\begin{figure}
\includegraphics[width=\textwidth,clip=true]{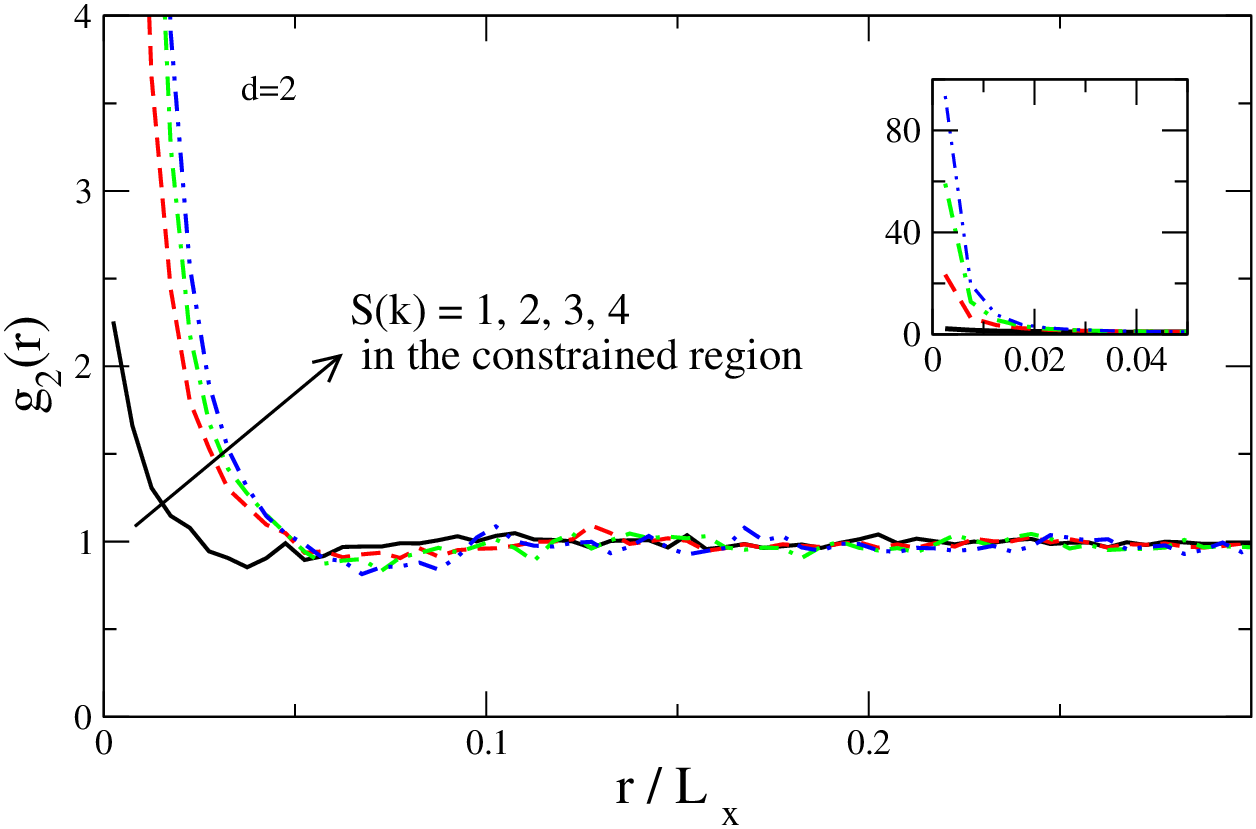}
\caption{Batten, Stillinger, Torquato}
\end{figure}

\clearpage
\newpage

\begin{figure}
\includegraphics[width=\textwidth,clip=true]{lumin2Dconfig.eps}
\centering
\caption{Batten, Stillinger, Torquato}
\end{figure}

\clearpage
\newpage

\begin{figure}
\includegraphics[width=\textwidth,clip=true]{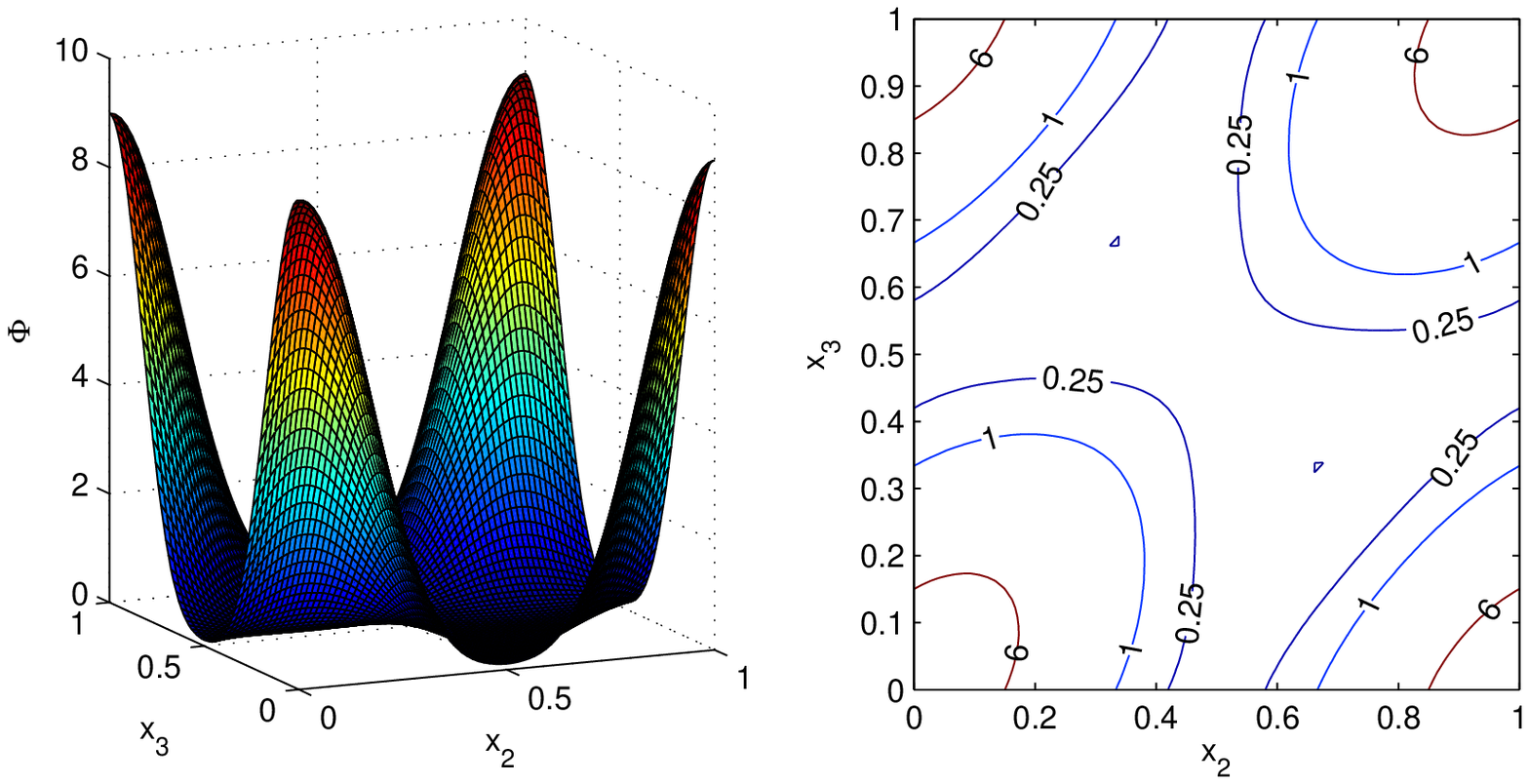}
\caption{Batten, Stillinger, Torquato}
\end{figure}

\clearpage
\newpage

\begin{figure}
\includegraphics[width=\textwidth,clip=true]{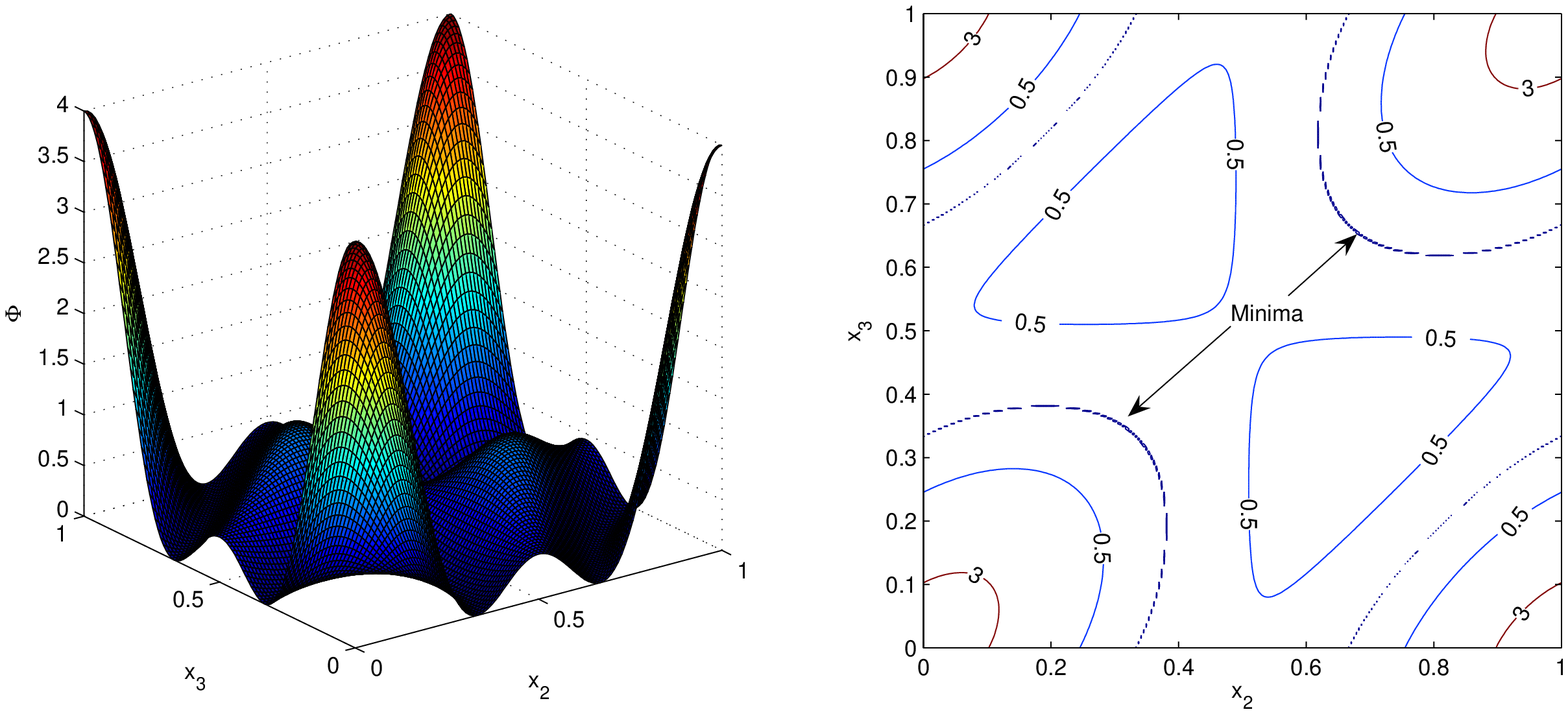}
\caption{Batten, Stillinger, Torquato}
\end{figure}


\begin{thebibliography}{48}
\expandafter\ifx\csname natexlab\endcsname\relax\def\natexlab#1{#1}\fi
\expandafter\ifx\csname bibnamefont\endcsname\relax
  \def\bibnamefont#1{#1}\fi
\expandafter\ifx\csname bibfnamefont\endcsname\relax
  \def\bibfnamefont#1{#1}\fi
\expandafter\ifx\csname citenamefont\endcsname\relax
  \def\citenamefont#1{#1}\fi
\expandafter\ifx\csname url\endcsname\relax
  \def\url#1{\texttt{#1}}\fi
\expandafter\ifx\csname urlprefix\endcsname\relax\def\urlprefix{URL }\fi
\providecommand{\bibinfo}[2]{#2}
\providecommand{\eprint}[2][]{\url{#2}}

\bibitem[{\citenamefont{Radin}(1987)}]{radin1987lta}
\bibinfo{author}{\bibfnamefont{C.}~\bibnamefont{Radin}},
  \bibinfo{journal}{Intl. J. Mod. Phys. B} \textbf{\bibinfo{volume}{1}},
  \bibinfo{pages}{1157} (\bibinfo{year}{1987}).

\bibitem[{\citenamefont{Torquato and Stillinger}(2006)}]{torquato2006ncl}
\bibinfo{author}{\bibfnamefont{S.}~\bibnamefont{Torquato}} \bibnamefont{and}
  \bibinfo{author}{\bibfnamefont{F.}~\bibnamefont{Stillinger}},
  \bibinfo{journal}{Exp. Math.} \textbf{\bibinfo{volume}{15}},
  \bibinfo{pages}{307} (\bibinfo{year}{2006}).

\bibitem[{\citenamefont{Fan et~al.}(1991)\citenamefont{Fan, Percus, Stillinger,
  and Stillinger}}]{fan1991ccd}
\bibinfo{author}{\bibfnamefont{Y.}~\bibnamefont{Fan}},
  \bibinfo{author}{\bibfnamefont{J.}~\bibnamefont{Percus}},
  \bibinfo{author}{\bibfnamefont{D.}~\bibnamefont{Stillinger}},
  \bibnamefont{and}
  \bibinfo{author}{\bibfnamefont{F.}~\bibnamefont{Stillinger}},
  \bibinfo{journal}{Phys. Rev. A} \textbf{\bibinfo{volume}{44}},
  \bibinfo{pages}{2394} (\bibinfo{year}{1991}).

\bibitem[{\citenamefont{Uche et~al.}(2004)\citenamefont{Uche, Stillinger, and
  Torquato}}]{uche2004ccd}
\bibinfo{author}{\bibfnamefont{O.}~\bibnamefont{Uche}},
  \bibinfo{author}{\bibfnamefont{F.}~\bibnamefont{Stillinger}},
  \bibnamefont{and} \bibinfo{author}{\bibfnamefont{S.}~\bibnamefont{Torquato}},
  \bibinfo{journal}{Phys. Rev. E} \textbf{\bibinfo{volume}{70}},
  \bibinfo{pages}{46122} (\bibinfo{year}{2004}).

\bibitem[{\citenamefont{Uche et~al.}(2006{\natexlab{a}})\citenamefont{Uche,
  Torquato, and Stillinger}}]{uche2006ccc}
\bibinfo{author}{\bibfnamefont{O.}~\bibnamefont{Uche}},
  \bibinfo{author}{\bibfnamefont{S.}~\bibnamefont{Torquato}}, \bibnamefont{and}
  \bibinfo{author}{\bibfnamefont{F.}~\bibnamefont{Stillinger}},
  \bibinfo{journal}{Phys. Rev. E} \textbf{\bibinfo{volume}{74}},
  \bibinfo{pages}{31104} (\bibinfo{year}{2006}{\natexlab{a}}).

\bibitem[{\citenamefont{Mattarelli et~al.}(2007)\citenamefont{Mattarelli,
  Montagna, and Verrocchio}}]{mattarelli2007ugc}
\bibinfo{author}{\bibfnamefont{M.}~\bibnamefont{Mattarelli}},
  \bibinfo{author}{\bibfnamefont{M.}~\bibnamefont{Montagna}}, \bibnamefont{and}
  \bibinfo{author}{\bibfnamefont{P.}~\bibnamefont{Verrocchio}},
  \bibinfo{journal}{Appl. Phys. Lett.} \textbf{\bibinfo{volume}{91}},
  \bibinfo{pages}{061911} (\bibinfo{year}{2007}).

\bibitem[{\citenamefont{Torquato et~al.}(2006)\citenamefont{Torquato, Uche, and
  Stillinger}}]{torquato2006rsa}
\bibinfo{author}{\bibfnamefont{S.}~\bibnamefont{Torquato}},
  \bibinfo{author}{\bibfnamefont{O.}~\bibnamefont{Uche}}, \bibnamefont{and}
  \bibinfo{author}{\bibfnamefont{F.}~\bibnamefont{Stillinger}},
  \bibinfo{journal}{Phys. Rev. E} \textbf{\bibinfo{volume}{74}},
  \bibinfo{pages}{61308} (\bibinfo{year}{2006}).

\bibitem[{\citenamefont{Joannopoulos et~al.}(1995)\citenamefont{Joannopoulos,
  Meade, and Winn}}]{joannopoulos1995pcm}
\bibinfo{author}{\bibfnamefont{J.}~\bibnamefont{Joannopoulos}},
  \bibinfo{author}{\bibfnamefont{R.}~\bibnamefont{Meade}}, \bibnamefont{and}
  \bibinfo{author}{\bibfnamefont{J.}~\bibnamefont{Winn}},
  \emph{\bibinfo{title}{{Photonic Crystals: Molding the Flow of Light}}}
  (\bibinfo{publisher}{Princeton University Press: Princeton, NJ},
  \bibinfo{year}{1995}).

\bibitem[{\citenamefont{Torquato}(2002)}]{torquato2002rhm}
\bibinfo{author}{\bibfnamefont{S.}~\bibnamefont{Torquato}},
  \emph{\bibinfo{title}{{Random Heterogeneous Materials: Microstucture and
  Macroscopic Properties}}} (\bibinfo{publisher}{SpringerVerlag: New York},
  \bibinfo{year}{2002}).

\bibitem[{foo()}]{footnote}
\bibinfo{note}{The conventional use of the term ``stealth" refers to any technology that
  creates invisibility to detection by radar, infrared, {\it etc.}, which can
  include altering material shape to deflect incident rays, using radiation
  absorbing materials or coatings, or using electronic jamming equipment. Here
  we use the term ``stealth material" only to refer to a material that is transparent to incident
  radiation.}

\bibitem[{\citenamefont{Schaefer}(1989)}]{schaefer1989pfa}
\bibinfo{author}{\bibfnamefont{D.}~\bibnamefont{Schaefer}},
  \bibinfo{journal}{Science} \textbf{\bibinfo{volume}{243}},
  \bibinfo{pages}{1023} (\bibinfo{year}{1989}).

\bibitem[{\citenamefont{Martin and Hurd}(1987)}]{martin1987sf}
\bibinfo{author}{\bibfnamefont{J.}~\bibnamefont{Martin}} \bibnamefont{and}
  \bibinfo{author}{\bibfnamefont{A.}~\bibnamefont{Hurd}}, \bibinfo{journal}{J.
  Appl. Cryst.} \textbf{\bibinfo{volume}{20}}, \bibinfo{pages}{61}
  (\bibinfo{year}{1987}).

\bibitem[{\citenamefont{Teixeira}(1988)}]{teixeira1988sas}
\bibinfo{author}{\bibfnamefont{J.}~\bibnamefont{Teixeira}},
  \bibinfo{journal}{J. Appl. Cryst.} \textbf{\bibinfo{volume}{21}},
  \bibinfo{pages}{781} (\bibinfo{year}{1988}).

\bibitem[{\citenamefont{Ashcroft and Mermin}(1976)}]{ashcroft1976ssp}
\bibinfo{author}{\bibfnamefont{N.}~\bibnamefont{Ashcroft}} \bibnamefont{and}
  \bibinfo{author}{\bibfnamefont{N.}~\bibnamefont{Mermin}},
  \emph{\bibinfo{title}{{Solid State Physics}}} (\bibinfo{publisher}{Saunders
  College: Philadelphia, Pa}, \bibinfo{year}{1976}).

\bibitem[{\citenamefont{Feynman}(1972)}]{feynman1998sm}
\bibinfo{author}{\bibfnamefont{R.}~\bibnamefont{Feynman}},
  \emph{\bibinfo{title}{{Statistical Mechanics}}}
  (\bibinfo{publisher}{Benjamin: Reading}, \bibinfo{year}{1972}).

\bibitem[{\citenamefont{Torquato and Stillinger}(2002)}]{torquato2002csr}
\bibinfo{author}{\bibfnamefont{S.}~\bibnamefont{Torquato}} \bibnamefont{and}
  \bibinfo{author}{\bibfnamefont{F.}~\bibnamefont{Stillinger}},
  \bibinfo{journal}{J. Phys. Chem. B} \textbf{\bibinfo{volume}{106}},
  \bibinfo{pages}{8354} (\bibinfo{year}{2002}).

\bibitem[{\citenamefont{Costin and Lebowitz}(2004)}]{costin2004cpd}
\bibinfo{author}{\bibfnamefont{O.}~\bibnamefont{Costin}} \bibnamefont{and}
  \bibinfo{author}{\bibfnamefont{J.}~\bibnamefont{Lebowitz}},
  \bibinfo{journal}{J. Phys. Chem. B} \textbf{\bibinfo{volume}{108}},
  \bibinfo{pages}{19614} (\bibinfo{year}{2004}).

\bibitem[{\citenamefont{Stillinger and Torquato}(2004)}]{stillinger2004pcf}
\bibinfo{author}{\bibfnamefont{F.}~\bibnamefont{Stillinger}} \bibnamefont{and}
  \bibinfo{author}{\bibfnamefont{S.}~\bibnamefont{Torquato}},
  \bibinfo{journal}{J. Phys. Chem. B} \textbf{\bibinfo{volume}{108}},
  \bibinfo{pages}{19589} (\bibinfo{year}{2004}).

\bibitem[{\citenamefont{Stillinger and Torquato}(2005)}]{stillinger2005rii}
\bibinfo{author}{\bibfnamefont{F.}~\bibnamefont{Stillinger}} \bibnamefont{and}
  \bibinfo{author}{\bibfnamefont{S.}~\bibnamefont{Torquato}},
  \bibinfo{journal}{Mol. Phys.} \textbf{\bibinfo{volume}{103}},
  \bibinfo{pages}{2943} (\bibinfo{year}{2005}).

\bibitem[{\citenamefont{Yeong and Torquato}(1998)}]{yeong1998rrm}
\bibinfo{author}{\bibfnamefont{C.}~\bibnamefont{Yeong}} \bibnamefont{and}
  \bibinfo{author}{\bibfnamefont{S.}~\bibnamefont{Torquato}},
  \bibinfo{journal}{Phys. Rev. E} \textbf{\bibinfo{volume}{57}},
  \bibinfo{pages}{495} (\bibinfo{year}{1998}).

\bibitem[{\citenamefont{Cule and Torquato}(1999)}]{cule1999grm}
\bibinfo{author}{\bibfnamefont{D.}~\bibnamefont{Cule}} \bibnamefont{and}
  \bibinfo{author}{\bibfnamefont{S.}~\bibnamefont{Torquato}},
  \bibinfo{journal}{J. Appl. Phys.} \textbf{\bibinfo{volume}{86}},
  \bibinfo{pages}{3428} (\bibinfo{year}{1999}).

\bibitem[{\citenamefont{Sheehan and Torquato}(2001)}]{sheehan2001gms}
\bibinfo{author}{\bibfnamefont{N.}~\bibnamefont{Sheehan}} \bibnamefont{and}
  \bibinfo{author}{\bibfnamefont{S.}~\bibnamefont{Torquato}},
  \bibinfo{journal}{J. Appl. Phys.} \textbf{\bibinfo{volume}{89}},
  \bibinfo{pages}{53} (\bibinfo{year}{2001}).

\bibitem[{\citenamefont{Crawford et~al.}(2003)\citenamefont{Crawford, Torquato,
  and Stillinger}}]{crawford2003acf}
\bibinfo{author}{\bibfnamefont{J.}~\bibnamefont{Crawford}},
  \bibinfo{author}{\bibfnamefont{S.}~\bibnamefont{Torquato}}, \bibnamefont{and}
  \bibinfo{author}{\bibfnamefont{F.}~\bibnamefont{Stillinger}},
  \bibinfo{journal}{J. Chem. Phys.} \textbf{\bibinfo{volume}{119}},
  \bibinfo{pages}{7065} (\bibinfo{year}{2003}).

\bibitem[{\citenamefont{Uche et~al.}(2006{\natexlab{b}})\citenamefont{Uche,
  Stillinger, and Torquato}}]{uche2006rpc}
\bibinfo{author}{\bibfnamefont{O.}~\bibnamefont{Uche}},
  \bibinfo{author}{\bibfnamefont{F.}~\bibnamefont{Stillinger}},
  \bibnamefont{and} \bibinfo{author}{\bibfnamefont{S.}~\bibnamefont{Torquato}},
  \bibinfo{journal}{Physica A} \textbf{\bibinfo{volume}{360}},
  \bibinfo{pages}{21} (\bibinfo{year}{2006}{\natexlab{b}}).

\bibitem[{\citenamefont{Ruelle}(1981)}]{ruelle1981mtf}
\bibinfo{author}{\bibfnamefont{D.}~\bibnamefont{Ruelle}}, \bibinfo{journal}{J.
  Stat. Phys.} \textbf{\bibinfo{volume}{26}}, \bibinfo{pages}{397}
  (\bibinfo{year}{1981}).

\bibitem[{\citenamefont{Ruelle}(1982)}]{ruelle1982tce}
\bibinfo{author}{\bibfnamefont{D.}~\bibnamefont{Ruelle}},
  \bibinfo{journal}{Physica A} \textbf{\bibinfo{volume}{113}},
  \bibinfo{pages}{619} (\bibinfo{year}{1982}).

\bibitem[{\citenamefont{Yukalov and Yukalova}(2001)}]{yukalov2001ptc}
\bibinfo{author}{\bibfnamefont{V.}~\bibnamefont{Yukalov}} \bibnamefont{and}
  \bibinfo{author}{\bibfnamefont{E.}~\bibnamefont{Yukalova}},
  \bibinfo{journal}{Int. J. Mod. Phys. B} \textbf{\bibinfo{volume}{15}},
  \bibinfo{pages}{2433} (\bibinfo{year}{2001}).

\bibitem[{\citenamefont{Butz et~al.}(1984)\citenamefont{Butz, Lerf, Saibene,
  and H{\"u}bler}}]{butz1984ers}
\bibinfo{author}{\bibfnamefont{T.}~\bibnamefont{Butz}},
  \bibinfo{author}{\bibfnamefont{A.}~\bibnamefont{Lerf}},
  \bibinfo{author}{\bibfnamefont{S.}~\bibnamefont{Saibene}}, \bibnamefont{and}
  \bibinfo{author}{\bibfnamefont{A.}~\bibnamefont{H{\"u}bler}},
  \bibinfo{journal}{Hyperfine Interactions} \textbf{\bibinfo{volume}{20}},
  \bibinfo{pages}{263} (\bibinfo{year}{1984}).

\bibitem[{\citenamefont{Radin}(1991)}]{radin1991dgs}
\bibinfo{author}{\bibfnamefont{C.}~\bibnamefont{Radin}}, \bibinfo{journal}{Rev.
  Math. Phys.} \textbf{\bibinfo{volume}{3}}, \bibinfo{pages}{125}
  (\bibinfo{year}{1991}).

\bibitem[{\citenamefont{Le~Berre et~al.}(2002)\citenamefont{Le~Berre, Ressayre,
  Tallet, Pomeau, and Di~Menza}}]{leberre2002ecc}
\bibinfo{author}{\bibfnamefont{M.}~\bibnamefont{Le~Berre}},
  \bibinfo{author}{\bibfnamefont{E.}~\bibnamefont{Ressayre}},
  \bibinfo{author}{\bibfnamefont{A.}~\bibnamefont{Tallet}},
  \bibinfo{author}{\bibfnamefont{Y.}~\bibnamefont{Pomeau}}, \bibnamefont{and}
  \bibinfo{author}{\bibfnamefont{L.}~\bibnamefont{Di~Menza}},
  \bibinfo{journal}{Phys. Rev. E} \textbf{\bibinfo{volume}{66}},
  \bibinfo{pages}{26203} (\bibinfo{year}{2002}).

\bibitem[{\citenamefont{Pusey et~al.}(1989)\citenamefont{Pusey, van Megen,
  Bartlett, Ackerson, Rarity, and Underwood}}]{pusey1989sch}
\bibinfo{author}{\bibfnamefont{P.}~\bibnamefont{Pusey}},
  \bibinfo{author}{\bibfnamefont{W.}~\bibnamefont{van Megen}},
  \bibinfo{author}{\bibfnamefont{P.}~\bibnamefont{Bartlett}},
  \bibinfo{author}{\bibfnamefont{B.}~\bibnamefont{Ackerson}},
  \bibinfo{author}{\bibfnamefont{J.}~\bibnamefont{Rarity}}, \bibnamefont{and}
  \bibinfo{author}{\bibfnamefont{S.}~\bibnamefont{Underwood}},
  \bibinfo{journal}{Phys. Rev. Lett.} \textbf{\bibinfo{volume}{63}},
  \bibinfo{pages}{2753} (\bibinfo{year}{1989}).

\bibitem[{\citenamefont{Torquato and Stillinger}(2003)}]{torquato2003ldf}
\bibinfo{author}{\bibfnamefont{S.}~\bibnamefont{Torquato}} \bibnamefont{and}
  \bibinfo{author}{\bibfnamefont{F.}~\bibnamefont{Stillinger}},
  \bibinfo{journal}{Phys. Rev. E} \textbf{\bibinfo{volume}{68}},
  \bibinfo{pages}{41113} (\bibinfo{year}{2003}).

\bibitem[{\citenamefont{Gabrielli et~al.}(2002)\citenamefont{Gabrielli, Joyce,
  and Sylos~Labini}}]{gabrielli2002glu}
\bibinfo{author}{\bibfnamefont{A.}~\bibnamefont{Gabrielli}},
  \bibinfo{author}{\bibfnamefont{M.}~\bibnamefont{Joyce}}, \bibnamefont{and}
  \bibinfo{author}{\bibfnamefont{F.}~\bibnamefont{Sylos~Labini}},
  \bibinfo{journal}{Phys. Rev. D} \textbf{\bibinfo{volume}{65}},
  \bibinfo{pages}{83523} (\bibinfo{year}{2002}).

\bibitem{Ka02}
A. R. Kansal, S. Torquato, and F. H. Stillinger, Phys. Rev. E {\bf 66}, 041109 (2002).

\bibitem[{\citenamefont{Donev et~al.}(2005)\citenamefont{Donev, Stillinger, and
  Torquato}}]{donev2005udf}
\bibinfo{author}{\bibfnamefont{A.}~\bibnamefont{Donev}},
  \bibinfo{author}{\bibfnamefont{F.}~\bibnamefont{Stillinger}},
  \bibnamefont{and} \bibinfo{author}{\bibfnamefont{S.}~\bibnamefont{Torquato}},
  \bibinfo{journal}{Phys. Rev. Lett.} \textbf{\bibinfo{volume}{95}},
  \bibinfo{pages}{90604} (\bibinfo{year}{2005}).

\bibitem[{\citenamefont{Gabrielli and Torquato}(2004)}]{gabrielli2004vav}
\bibinfo{author}{\bibfnamefont{A.}~\bibnamefont{Gabrielli}} \bibnamefont{and}
  \bibinfo{author}{\bibfnamefont{S.}~\bibnamefont{Torquato}},
  \bibinfo{journal}{Phys. Rev. E} \textbf{\bibinfo{volume}{70}},
  \bibinfo{pages}{41105} (\bibinfo{year}{2004}):
  \bibinfo{author}{\bibfnamefont{A.}~\bibnamefont{Gabrielli}}, 
  \bibinfo{author}{\bibfnamefont{M.}~\bibnamefont{Joyce}}, \bibnamefont{and}
  \bibinfo{author}{\bibfnamefont{S.}~\bibnamefont{Torquato}},
  \bibinfo{journal}{Phys. Rev. E}  (\bibinfo{year}{In press}).
  
\bibitem[{\citenamefont{Press}(1992)}]{press1992nrc}
\bibinfo{author}{\bibfnamefont{W.}~\bibnamefont{Press}},
  \emph{\bibinfo{title}{{Numerical Recipes in C: The Art of Scientific
  Computing}}} (\bibinfo{publisher}{Cambridge University Press},
  \bibinfo{year}{1992}).

\bibitem[{\citenamefont{Dennis and Mei}(1979)}]{dennis1979tnu}
\bibinfo{author}{\bibfnamefont{J.}~\bibnamefont{Dennis}} \bibnamefont{and}
  \bibinfo{author}{\bibfnamefont{H.}~\bibnamefont{Mei}}, \bibinfo{journal}{J.
  Optim. Theory Appl.} \textbf{\bibinfo{volume}{28}}, \bibinfo{pages}{453}
  (\bibinfo{year}{1979}).

\bibitem[{\citenamefont{Kaufman}(1999)}]{kaufman1999rsq}
\bibinfo{author}{\bibfnamefont{L.}~\bibnamefont{Kaufman}},
  \bibinfo{journal}{SIAM J. Optim.} \textbf{\bibinfo{volume}{10}},
  \bibinfo{pages}{56} (\bibinfo{year}{1999}).

\bibitem[{\citenamefont{MacGillivray and Dodd}(1986)}]{macgillivray1986fso}
\bibinfo{author}{\bibfnamefont{H.}~\bibnamefont{MacGillivray}}
  \bibnamefont{and} \bibinfo{author}{\bibfnamefont{R.}~\bibnamefont{Dodd}},
  \bibinfo{journal}{J. Astrophys. and Astronomy} \textbf{\bibinfo{volume}{7}},
  \bibinfo{pages}{293} (\bibinfo{year}{1986}).

\bibitem[{\citenamefont{S{\"u}t{\H{o}}}(2005)}]{suto2005cgs}
\bibinfo{author}{\bibfnamefont{A.}~\bibnamefont{S{\"u}t{\H{o}}}},
  \bibinfo{journal}{Phys. Rev. Lett.} \textbf{\bibinfo{volume}{95}},
  \bibinfo{pages}{265501} (\bibinfo{year}{2005}).

\bibitem[{\citenamefont{Torquato and Stillinger}(2008)}]{torquato2008ndr}
\bibinfo{author}{\bibfnamefont{S.}~\bibnamefont{Torquato}} \bibnamefont{and}
  \bibinfo{author}{\bibfnamefont{F.}~\bibnamefont{Stillinger}},
  \bibinfo{journal}{Phys. Rev. Lett.} \textbf{\bibinfo{volume}{100}},
  \bibinfo{pages}{020602} (\bibinfo{year}{2008}).

\bibitem[{\citenamefont{John}(1987)}]{john1987slp}
\bibinfo{author}{\bibfnamefont{S.}~\bibnamefont{John}}, \bibinfo{journal}{Phys.
  Rev. Lett.} \textbf{\bibinfo{volume}{58}}, \bibinfo{pages}{2486}
  (\bibinfo{year}{1987}).

\bibitem[{\citenamefont{Yablonovitch}(1987)}]{yablonovitch1987ise}
\bibinfo{author}{\bibfnamefont{E.}~\bibnamefont{Yablonovitch}},
  \bibinfo{journal}{Phys. Rev. Lett.} \textbf{\bibinfo{volume}{58}},
  \bibinfo{pages}{2059} (\bibinfo{year}{1987}).

\bibitem[{\citenamefont{Yablonovitch}(1993)}]{yablonovitch1993pbg}
\bibinfo{author}{\bibfnamefont{E.}~\bibnamefont{Yablonovitch}},
  \bibinfo{journal}{J. Opt. Soc. Am. B} \textbf{\bibinfo{volume}{10}},
  \bibinfo{pages}{283} (\bibinfo{year}{1993}).

\bibitem[{\citenamefont{Zhang et~al.}(2000)\citenamefont{Zhang, Lei, Wang,
  Zheng, Tam, Chan, and Sheng}}]{zhang2000rpb}
\bibinfo{author}{\bibfnamefont{W.}~\bibnamefont{Zhang}},
  \bibinfo{author}{\bibfnamefont{X.}~\bibnamefont{Lei}},
  \bibinfo{author}{\bibfnamefont{Z.}~\bibnamefont{Wang}},
  \bibinfo{author}{\bibfnamefont{D.}~\bibnamefont{Zheng}},
  \bibinfo{author}{\bibfnamefont{W.}~\bibnamefont{Tam}},
  \bibinfo{author}{\bibfnamefont{C.}~\bibnamefont{Chan}}, \bibnamefont{and}
  \bibinfo{author}{\bibfnamefont{P.}~\bibnamefont{Sheng}},
  \bibinfo{journal}{Phys. Rev. Lett.} \textbf{\bibinfo{volume}{84}},
  \bibinfo{pages}{2853} (\bibinfo{year}{2000}).

\bibitem[{\citenamefont{Asatryan et~al.}(1999)\citenamefont{Asatryan, Robinson,
  Botten, McPhedran, Nicorovici, and Martijn~de Sterke}}]{asatryan1999edw}
\bibinfo{author}{\bibfnamefont{A.}~\bibnamefont{Asatryan}},
  \bibinfo{author}{\bibfnamefont{P.}~\bibnamefont{Robinson}},
  \bibinfo{author}{\bibfnamefont{L.}~\bibnamefont{Botten}},
  \bibinfo{author}{\bibfnamefont{R.}~\bibnamefont{McPhedran}},
  \bibinfo{author}{\bibfnamefont{N.}~\bibnamefont{Nicorovici}},
  \bibnamefont{and} \bibinfo{author}{\bibfnamefont{C.}~\bibnamefont{Martijn~de
  Sterke}}, \bibinfo{journal}{Phys. Rev. E} \textbf{\bibinfo{volume}{60}},
  \bibinfo{pages}{6118} (\bibinfo{year}{1999}).

\bibitem[{\citenamefont{Johnson et~al.}(2002)\citenamefont{Johnson, Ibanescu,
  Skorobogatiy, Weisberg, Joannopoulos, and Fink}}]{johnson2002ptm}
\bibinfo{author}{\bibfnamefont{S.}~\bibnamefont{Johnson}},
  \bibinfo{author}{\bibfnamefont{M.}~\bibnamefont{Ibanescu}},
  \bibinfo{author}{\bibfnamefont{M.}~\bibnamefont{Skorobogatiy}},
  \bibinfo{author}{\bibfnamefont{O.}~\bibnamefont{Weisberg}},
  \bibinfo{author}{\bibfnamefont{J.}~\bibnamefont{Joannopoulos}},
  \bibnamefont{and} \bibinfo{author}{\bibfnamefont{Y.}~\bibnamefont{Fink}},
  \bibinfo{journal}{Phys. Rev. E} \textbf{\bibinfo{volume}{65}},
  \bibinfo{pages}{66611} (\bibinfo{year}{2002}).

\bibitem[{\citenamefont{Hansen and McDonald}(2006)}]{hansen2006tsl}
\bibinfo{author}{\bibfnamefont{J.}~\bibnamefont{Hansen}} \bibnamefont{and}
  \bibinfo{author}{\bibfnamefont{I.}~\bibnamefont{McDonald}},
  \emph{\bibinfo{title}{{Theory of Simple Liquids}}}
  (\bibinfo{publisher}{Academic Press: New York}, \bibinfo{year}{2006}).

\end{thebibliography}
\end{document}